\documentclass[english,aps,prl,amsmath,amssymb,twocolumn,letterpaper,citeautoscript,superscriptaddress,longbibliography]{revtex4-1}

\usepackage{mathtools} 

\usepackage[normalem]{ulem} 

\usepackage{comment}
\usepackage{amsmath}
\usepackage{physics}
\usepackage{graphicx}
\usepackage{babel}
\usepackage{amstext}
\usepackage{esint}
\usepackage{multirow}
\usepackage[bookmarks=false,linkcolor=blue,urlcolor=blue,colorlinks,citecolor=blue]{hyperref}
\usepackage{pdfpages}	
\usepackage{pgffor}		

\makeatletter
\AtBeginDocument{\let\LS@rot\@undefined} 
\makeatother							 

\begin{document}

\title{Edge spin transport in the disordered two-dimensional topological insulator WTe$_2$}

\author{Justin Copenhaver}

\affiliation{Department of Physics and Astronomy, Purdue University, West Lafayette, Indiana 47907 USA}

\author{Jukka I. V\"ayrynen}

\affiliation{Department of Physics and Astronomy, Purdue University, West Lafayette, Indiana 47907 USA}
\affiliation{Purdue Quantum Science and Engineering Institute, Purdue University, West Lafayette, Indiana 47907, USA}

\date{\today}
\begin{abstract} 
The spin conductance of two-dimensional topological insulators (2D TIs) is not expected to be quantized in the presence of perturbations that break the spin-rotational symmetry. However, the deviation from the pristine-limit quantization has yet to be studied in detail. In this paper, we define the spin current operator for the helical edge modes of a 2D TI and introduce a four-terminal setup to measure spin conductances. Using the developed formalism, we consider the effects of disorder terms that break spin-rotational symmetry or give rise to edge-to-edge coupling. We identify a key role played by spin torque in an out-of-equilibrium edge. We then utilize a tight-binding model of topological monolayer WTe$_2$ and scattering matrix formalism to numerically study spin transport in a four-terminal 2D TI device. In particular, we calculate the spin conductances and characteristic spin decay length in the presence of magnetic disorder. In addition, we study the effects of inter-edge scattering in a quantum point contact geometry. We find that the spin Hall conductance is surprisingly robust to spin symmetry-breaking perturbations, as long as time-reversal symmetry is preserved and inter-edge scattering is weak. 
\end{abstract}
\maketitle

Electrical control of spins is one of the central objectives in the field of spintronics~\cite{RevModPhys.76.323}. Topological insulators (TIs) are materials with strong spin-orbit coupling and host  spin-momentum locked gapless modes confined to the boundary of an insulating bulk~\cite{hasanRMP,qi_review}. These helical boundary modes offer new possibilities to generate spin polarization and spin currents with electrical means~\cite{2012NatMa..11..409P,he2019topological,culcer2020transport}. So far, most studies of topological insulators from a spintronics point of view have focused on 3D TIs~\cite{2014NatMa..13..699F,2014Natur.511..449M,tang2014electrical,Tian2015}, whose 2D surface  hosts a massless helical Dirac fermion.
(This surface  is somewhat similar to graphene, which hosts two Dirac cones and has also been subject to extensive spintronics research~\cite{grapheneReview2014,Roche_2015}.)

However, impurity scattering limits the potential of using  the 3D TI surface states for spintronics. Even though direct backscattering $\mathbf{k} \to -\mathbf{k}$ of the Dirac electrons is forbidden by time-reversal symmetry (since $\mathbf{k}$ and $ -\mathbf{k}$ are oppositely spin-polarized), scattering by any other angle is allowed, which leads to the loss of momentum and spin conservation at a scale set by the elastic mean free path~\cite{2012NatMa..11..409P}. By the same token, current-induced spin accumulation is similarly limited by the mean free path~\cite{PhysRevB.82.155457}.

Impurity scattering is much more restricted in 2D TIs whose boundary modes are confined to 1D. These helical modes have only 2 momentum directions, left and right, and time-reversal symmetry (TRS) forbids elastic backscattering between the two. The modes therefore remain ballistic (and retain their spin) at distances below the inelastic mean free path~\cite{wu06,PhysRevB.73.045322,schmidt12,budich12,Vayrynen2014,2014PhRvB..90g5118K,PhysRevLett.121.106601,mcginley2020fragility}. Likewise, current-induced out-of-equilibrium spin polarization of a 2D TI edge is not limited by elastic non-magnetic impurity scattering. Indeed, a bias voltage $V$ (or charge current $e^2V/h$) leads to a spin accumulation per density $ \langle S_z \rangle / n = eV/(4 E_F)$ on a 2D TI edge, independent of scalar disorder (the opposite edge would have the opposite spin polarization). Here we denote $\mathbf{z}$ the spin quantization axis at the Fermi level, assuming it does not vary on the scale $eV$. 

Spin transport on the one-dimensional edge states of a 2D TI was first considered in Refs.~\cite{kane_quantum_2005,PhysRevLett.96.106802} where the spin Hall conductance was calculated in the ideal case with the conservation of spin-$z$ projection. In this case, the spin Hall conductance is found to be quantized to $e/(4\pi)$. Upon breaking the spin conservation, the spin Hall conductance is generally finite but not expected to be quantized~\cite{PhysRevLett.95.146802,PhysRevLett.95.136602,PhysRevB.74.085308}. 

Various spin-rotation symmetry breaking  mechanisms on the 2D TI edge have been considered in the context of charge transport~\cite{wu06,PhysRevB.73.045322,tanaka11,budich12,2012PhRvB..85w5304L,schmidt12,altshuler13,Vayrynen2014,2014PhRvB..90g5118K,2019PhRvL.122a6601N}. On a clean, translationally invariant edge, the spin rotational symmetry may be broken due to bulk or structural inversion asymmetry which can lead to a momentum space spin rotation of the helical edge modes~\cite{schmidt12,2015PhRvB..91x5112R}, without breaking  time-reversal symmetry. Similarly, the spin quantization axis may rotate in real space in the presence of a random Rashba spin-orbit term~\cite{PhysRevLett.116.086603,strom10,PhysRevB.102.035423}. These TRS mechanisms do not lead to elastic backscattering but can modify the charge conductance at non-zero temperatures  inelastically~\cite{wu06,PhysRevB.73.045322,schmidt12,2012PhRvB..85w5304L}. Elastic backscattering becomes possible when TRS is broken~\cite{wu06,PhysRevB.73.045322,2019PhRvL.122a6601N}. This can be achieved for example by applying an external magnetic field~\cite{ma2015unexpected,2017NatPh..13..677F,2018Sci...359...76W,PhysRevB.102.161402,2021arXiv211005718S} or by doping the sample with polarized magnetic impurities~\cite{Jack2020,shamim2021quantized} which both suppress edge conduction. While spin-non-conserving perturbations have received considerable attention in charge transport, relatively few quantitative studies~\cite{PhysRevLett.95.136602,PhysRevB.93.165414,PhysRevLett.122.196601,wu2020computational,2020arXiv200705626G} have focused on spin transport in 2D TIs. 

In this paper we formulate the low-energy scattering theory of spin transport in 2D TI edge states and use numerical simulations to go beyond the effective model. Focusing on the recently discovered monolayer WTe$_2$ topological insulator~\cite{qian2014quantum,2017NatPh..13..677F,2018Sci...359...76W,2017NatPh..13..683T,2017PhRvB..96d1108J,2017NatCo...8..659P,Lau2019} as an example, we carry out an extensive numerical study of disorder effects on   spin transport. We consider both spin-conserving and explicitly spin-symmetry-breaking terms such as random scalar on-site disorder, spin-non-conserving disorder in the spin-orbit coupling strength, TRS breaking magnetic impurities, as well as inter-edge scattering in a quantum point contact geometry.

Our analytical theory clarifies how the spin conductance quantization gets broken by spin non-conserving perturbations. We identify a crucial role played by local equilibrium or non-equilibrium on the TI edge. Namely, the non-conservation of edge spin current (and a resulting non-quantized spin conductance) arises from a spin torque generated by the spin non-conserving disorder. As we will show, the spin torque vanishes if the edge is in local equilibrium, and is generally non-zero when the edge is out of equilibrium (and can have a non-zero $\langle S_z \rangle$). As a result, when using a 4-terminal measurement of the spin conductances, the bias configuration is of key importance: when the edge has no voltage drop, it can carry a conserved spin current, see Figs.~\ref{Conductance setups}--\ref{Four-terminal device} and Table~\ref{Conductance definitions table}.    

The outline of our paper is as follows. We first introduce an effective 1D model for the helical edge modes (Sec.~\ref{Effective Hamiltonian Sec}). We derive the spin current operator and discuss how  intra- and inter-edge backscattering perturbations modify the average spin current. In Sec.~\ref{Current Sec}, we introduce the spin-resolved Landauer-B\"uttiker formula to define the spin conductances for a multiterminal setup. In Sec.~\ref{Results Sec}, we present our numerical simulations for spin transport in disordered multiterminal systems and in Sec.~\ref{Conclusions Sec} we draw our conclusions.

\section{\label{Effective Hamiltonian Sec}Effective description  of edge spin transport}

In this section we develop a low-energy effective Hamiltonian which describes the propagation of the helical edge states in a 2D TI. We then utilize this model to study the effects of localized magnetic disorder and inter-edge scattering on the spin transport properties of the material. 

The characteristic feature of a 2D TI is the presence of a pair of helical edge modes and a gapped bulk. On a given edge and at a fixed energy, the helical modes have opposite spin-polarizations and velocities. At low energies, we can approximate the edge spectrum by a linear dispersion and ignore any momentum space spin rotation~\cite{2015PhRvB..91x5112R}. Denoting $\mathbf{z}$ the spin quantization axis of the TI, we obtain the 1D effective Hamiltonian of a single edge,
\begin{equation}
    H_0 = \int dx\, \Psi^\dagger (-i\hbar v \partial_x\sigma_z - \mu) \Psi \,,
    \label{Effective Hamiltonian}
\end{equation}
where $v$ is the  velocity of the edge modes, $\mu$ is the chemical potential, $\sigma_{i}$ denotes the spin Pauli matrices, and $\Psi(x) = \left(\psi_\uparrow, \psi_\downarrow\right)^T$ is the electron field operator.

While the effective Hamiltonian~\eqref{Effective Hamiltonian} does not have full spin-rotational symmetry, it does have a $U(1)$ spin-rotational symmetry about the $\mathbf{z}$-axis; we can therefore define a conserved spin current along this axis. Starting from the spin density $S_z(x) = \frac{\hbar}{2} \Psi^\dagger(x) \sigma_z \Psi(x)$, we  obtain the spin-$z$ current operator by using the continuity equation~\footnote{The eigenstates of Eq.~(\ref{Effective Hamiltonian}) carry no spin current along the $x$ or $y$ axes.}\cite{Shi2006, Marcelli2020}:
\begin{equation}
    \partial_t S_z + \partial_x I^s_z = 0 \,.
    \label{Continuity equation}
\end{equation}
The time derivative in Eq.~(\ref{Continuity equation}) can be evaluated using the Heisenberg equation of motion: $\partial_t S_z = \frac{i}{\hbar}\comm{H_0}{S_z}$. The commutator can then be expressed in terms of the gradient of the density operator $\rho(x) = \Psi^\dagger(x) \Psi(x)$. Remarkably, the spin current along the conserved axis is thus tied to the local density:
\begin{equation}
    I^s_z = \frac{\hbar v}{2}\rho \,.
    \label{Spin current analytical}
\end{equation}
This simple result is a direct consequence of spin-momentum locking: left and right moving electrons carry equal spin currents since they have opposite velocities \textit{and} spin projections~\footnote{We note in passing that the spin current, Eq.~(\ref{Spin current analytical}), is proportional to the conserved number density while  the charge current  is proportional to the conserved spin density $S_z$.}. This is in contrast to conduction by spin degenerate states that are not spin-momentum locked and carry no net spin current.

\begin{figure*}
    \centering
    \includegraphics[width=.32\textwidth]{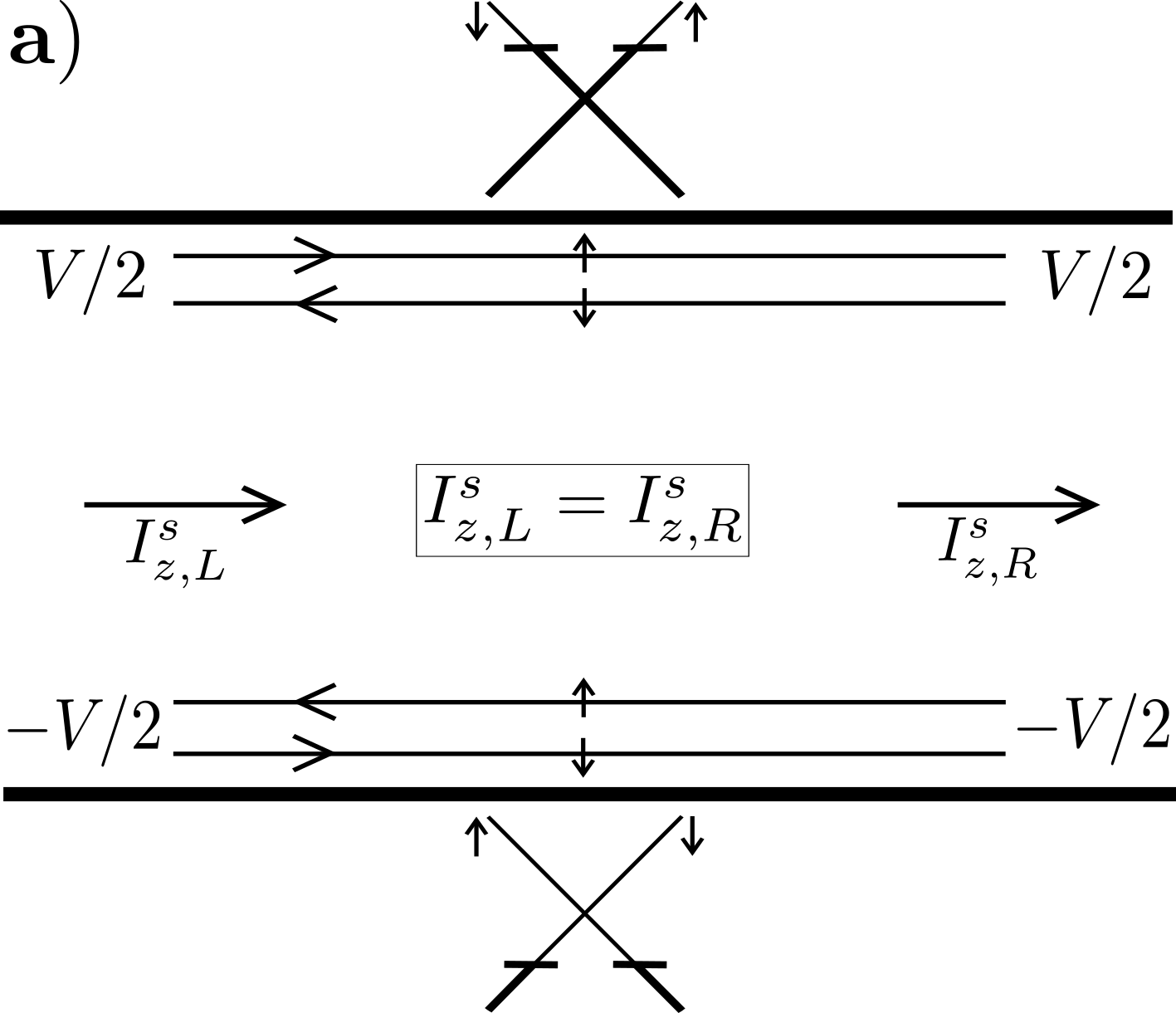} 
    \includegraphics[width=.32\textwidth]{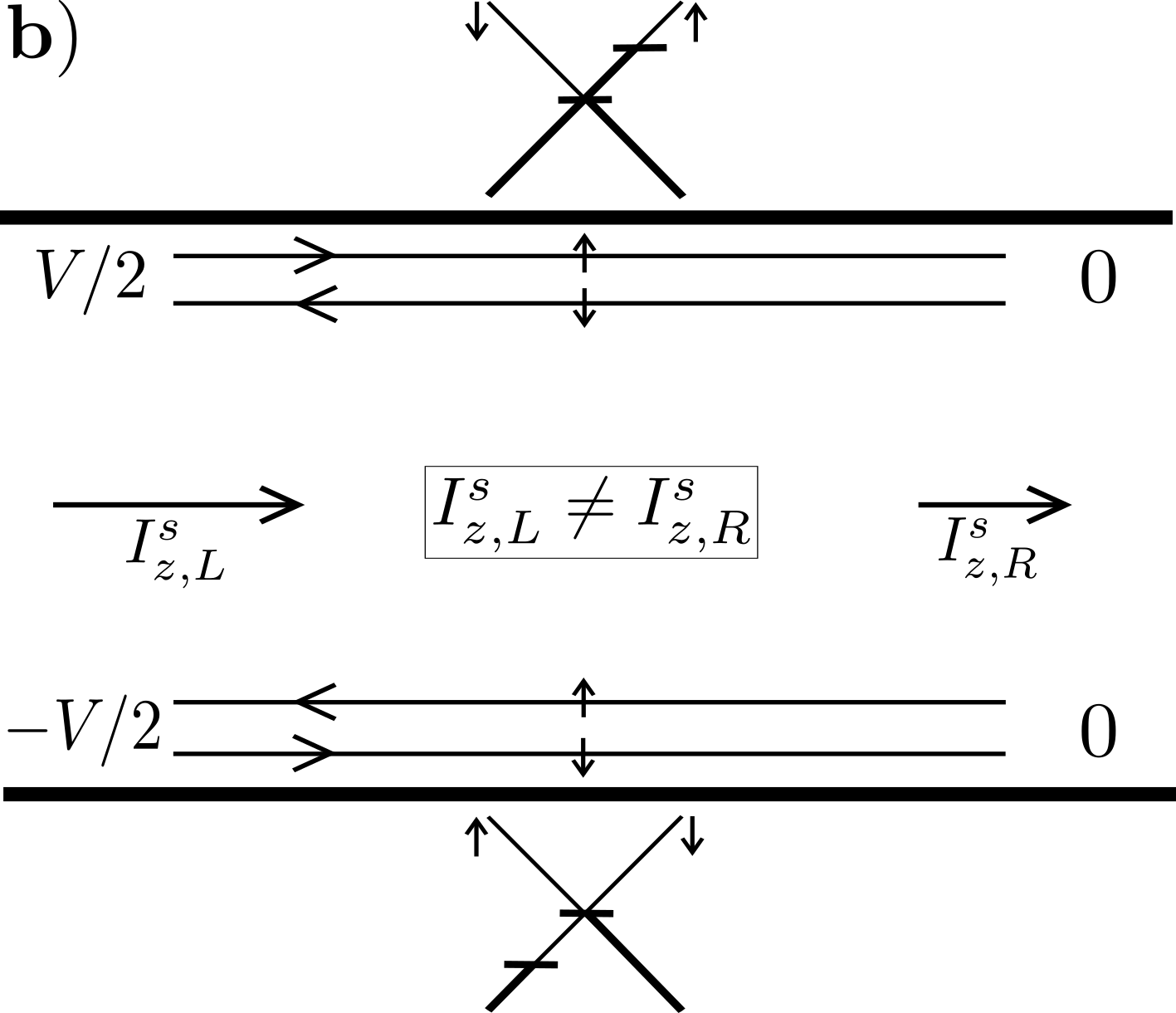}
    \includegraphics[width=.32\textwidth]{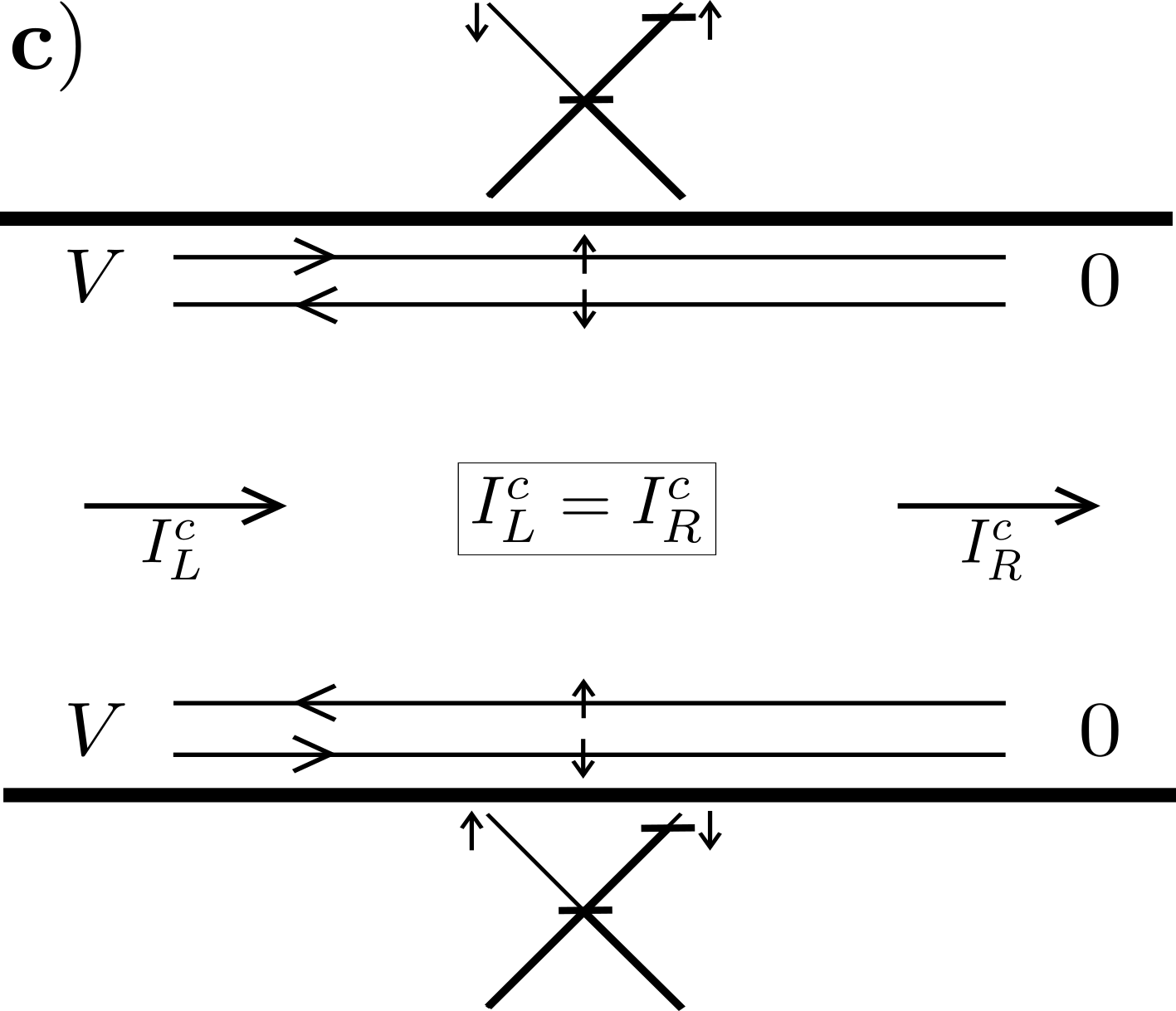}
    \caption{Schematic diagrams demonstrating the  voltage setups considered in this paper. Each diagram corresponds to a particular voltage arrangement indicated by the corner values. The edge states along with their propagation directions and spin orientations are depicted by the solid lines; the Dirac cones indicate the filling of these states. The relation between the left and right currents in the presence of $S_z$ non-conserving disorder can be deduced using the transmission and reflection coefficients from Eqs.~(\ref{Transmission coefficient})--(\ref{Reflection coefficient}) and is shown in the center of each diagram.
    \textbf{a)} Voltage setup of a 2D TI nanoribbon with a quantized spin Hall conductance $G^s_H = I^s_{z,L}/V = I^s_{z,R}/V$. Only perturbations which cause bulk conduction or couple the top and bottom edges will cause a deviation from the quantized conductance value. In the absence of such perturbations, the spin current is conserved since each edge is at a local equilibrium and spin torque vanishes. \textbf{b)} Voltage setup producing non-quantized spin conductances when $S_z$ non-conserving disorder is present. Due to the non-equilibrium distribution on each edge, there is a non-zero spin torque which breaks the conservation of spin current, $I^s_{z,L} \neq I^s_{z,R} $. The lack of spin current conservation requires the definition of separate incident and transmitted spin conductances given by $G^s_I = I^s_{z,L}/V$ and $G^s_T = I^s_{z,R}/V$, respectively. \textbf{c)} Standard setup used to define the two-terminal charge conductance $G^c_{2T} = I^c_L/V = I^c_R/V$. Here, voltage distribution of each edge is the same, resulting in no net horizontal spin current.}
    \label{Conductance setups}
\end{figure*}

Importantly, we note that any local perturbation which does not break the $U(1)$ spin symmetry  of Eq.~(\ref{Effective Hamiltonian}) will not modify the spin current. We will see below that the spin current is indeed robust against such perturbations. One might expect even greater robustness of the spin current since  $I^s_z$, Eq.~(\ref{Spin current analytical}), commutes with any particle number conserving operator. This robustness is manifest in the quantization of the spin Hall conductance of a two-edge system, as long as inter-edge scattering (which breaks the conservation of particle number on a given edge) is absent and each edge is at a local equilibrium, see Fig.~\ref{Conductance setups}a. However, random spin-orbit coupling or magnetic disorder terms $\delta H$ in the Hamiltonian can break the $S_z$ conservation, leading to a spin-torque term on the right-hand-side of Eq.~(\ref{Continuity equation}),
\begin{equation}
    \mathcal{T} = -\frac{i}{\hbar}\comm{\delta H}{S_z}  \,.
    \label{eq:SpinTorque}
\end{equation}
In general, this  spin torque breaks the conservation of the spin current defined by Eq.~(\ref{Spin current analytical})~\footnote{One could in principle solve the continuity equation with the spin torque $\mathcal{T}$  absorbed into the divergence of the current; the spin current defined from such equation would then be conserved~\cite{Shi2006, Marcelli2020}. One would then need to project the spin current found in this way onto the $z$-axis to determine the spin-$z$ current. As $\mathcal{T}$ is not easy to evaluate in general, however, a simpler solution is to stick to the definition from Eq.~(\ref{Spin current analytical}) and allow for the possibility of current sources/sinks where spin torque is present.}. We will see that in an out-of-equilibrium situation the spin torque can be on average non-zero and lead to a deviation of the spin conductance from the quantized value, see Fig.~\ref{Conductance setups}b.

To study the effect of $S_z$-non-conserving magnetic perturbations, we begin by adding a spatially-dependent disorder term to Eq.~(\ref{Effective Hamiltonian}):
\begin{equation}
     \delta H = \int dx\, m(x) \Psi^\dagger \sigma_x \Psi \,.
    \label{Effective Hamiltonian disorder}
\end{equation}
The $\sigma_x$ operator in Eq.~(\ref{Effective Hamiltonian disorder}) breaks time-reversal (TR) symmetry and the $U(1)$ spin-symmetry, coupling right- and left-movers and resulting in spin-flipping reflections. We will assume  that $m(x)$ is non-zero only in the region between $0$ and $x_0$ so that we may  treat the system as a scattering problem. 

In the presence of the magnetic disorder, the spin torque term,  Eq.~(\ref{eq:SpinTorque}), is non-zero. Thus, the spin current as defined in Eq.~(\ref{Spin current analytical}) is no longer conserved in the disordered region. This leads to a discontinuity in the current due to the perturbation:
\begin{equation}
    I^s_z(x_0) - I^s_z(0) = \int_0^{x_0} dx\, \mathcal{T} = -\int_0^{x_0} dx\, m(x) \Psi^\dagger \sigma_y \Psi \,.
    \label{Discontinuity}
\end{equation}

This discontinuity can  be evaluated explicitly by using the scattering matrix to calculate the spin current in the left and right regions due to, say, an incident right-mover with unit amplitude. The  transmission  and reflection  coefficients $t$ and $r$ corresponding to Eq.~(\ref{Effective Hamiltonian disorder}) are given by  (see Appendix~\ref{Trans/Refl derivation})
\begin{align}
     t &= \sech{\eta_m} \,, \label{Transmission coefficient} \\
     r &= -i\tanh{\eta_m} \,, \label{Reflection coefficient}
\end{align}
where $\eta_m = \int^{x_0}_0 m(x) \,dx/(\hbar v)$ and we neglect the energy-dependence of the scattering amplitudes (assuming scattering states near the Dirac point). We can then use the scattering matrix $\mathbb{S}$ to relate the coefficients of the incoming modes $\Psi_{in}$ to the outgoing modes $\Psi_{out}$ by $\Psi_{out} = \mathbb{S}\Psi_{in}$, where
\begin{equation}
    \mathbb{S} = \begin{pmatrix}
        r & t \\
        t & r
    \end{pmatrix} \,.
    \label{Scattering matrix}
\end{equation}
For our incident right-mover of unit amplitude, the spin current in the left ($x<0$) and right ($x>x_0$) regions are related to the transmission and reflection coefficients by
\begin{align}
    &I^s_z(0) = \frac{\hbar v}{2}\left(1 + \abs{r}^2\right) = \frac{\hbar v}{2}\left(1 + \tanh^2{\eta_m}\right) \,, \label{Magnetic disorder left current} \\
    &I^s_z(x_0) = \frac{\hbar v}{2}\abs{t}^2 = \frac{\hbar v}{2}\left(1 - \tanh^2{\eta_m}\right) \,. \label{Magnetic disorder right current}
\end{align}
We see that the jump, or loss, in the spin current is then $I^s_z(x_0) - I^s_z(0) = -\hbar v \tanh^2{\eta_m}$.

We note that for large $\eta_m$, the ``transmitted'' spin current $I^s_z(x_0)$ becomes exponentially small, i.e.
\begin{equation}
    I^s_z(x_0) \approx 2\hbar v e^{-x_0 /l_0} \,,
    \label{eq:IsTExponential}
\end{equation}
where $l_0 = x_0/(2\eta_m )$ is a characteristic spin decay length. The transmitted spin current therefore decreases in the same way that transmitted charge current (and conductance) would.

The analysis leading to Eqs.~(\ref{Magnetic disorder left current})--(\ref{Magnetic disorder right current}) applied to an incident left-mover from the right shows spin currents with the values of $I^s_z(0)$ and $I^s_z(x_0)$ interchanged, i.e., a spin current $I^s_z(0)$, Eq.~(\ref{Magnetic disorder left current}), on the right of the barrier.  Hence, in general spin-flipping reflections lead to an increase in the spin current on the incident side and a decrease of equal magnitude on the transmitted side. 
In particular, when edge modes are incident with the same amplitude from both sides, the spin current per unit momentum is equal on both sides of the barrier, $I^s_z(0) = I^s_z(x_0) = \hbar v$, independent of the strength of spin-flip scattering. In this case the spin torque, Eq.~(\ref{Discontinuity}), vanishes; the magnetic impurities experience no spin torque in equilibrium~\cite{Vayrynen2016a}. This is a key observation that leads to the robustness of the spin Hall conductance in a four-terminal system when the edge is in local equilibrium, as will be discussed below.

Above, we evaluated the spin current carried by a single scattering state on a helical edge. The thermally averaged spin current for a single edge [obtained by averaging Eq.~(\ref{Spin current analytical})] is not mathematically well-defined (without a UV cutoff) nor physical. In an actual two-terminal device, there are two edges carrying opposite spin currents, which ensures that the total spin current vanishes at equilibrium. The single-edge Hamiltonian of Eq.~(\ref{Effective Hamiltonian}) can be  extended to include both edges of a 2D TI ribbon by introducing another set of  Pauli matrices $\tau_i$ that act on the edge degree of freedom. The effective Hamiltonian of two uncoupled edges at the same chemical potential $\mu$ is given by
\begin{equation}
    H_0 = \int dx\, \tilde{\Psi}^\dagger (-i\hbar v \partial_x\sigma_z \tau_z - \mu) \tilde{\Psi} \,,
    \label{Effective Hamiltonian two edges}
\end{equation}
where $\tilde{\Psi} = (\Psi_1, \Psi_2 )^T $ denotes the two-edge field operator and $\Psi_i = ( \psi_{i,\uparrow}, \psi_{i,\downarrow})$. The matrix $\tau_z$ in the kinetic energy term ensures that the two edges carry edge modes with opposite helicities. Generalizing Eq.~(\ref{Spin current analytical}) to the two-edge system, we obtain the spin current operator
\begin{equation}
    I^s_z = \frac{\hbar v}{2} (\rho_1 - \rho_2) \,.
    \label{Spin current two edges}
\end{equation}
which consists of counter-propagating spin currents on the two edges 1 and 2.

A spin Hall current can  be driven if the two edges  of the ribbon are held at different, constant chemical potentials. This can be modeled by setting $\mu \to \mu + \tau_z eV/2$ in Eq.~(\ref{Effective Hamiltonian two edges}). Such an inter-edge bias can be achieved, for example, by using four terminals (see Fig.~\ref{Conductance setups}a and Sec.~\ref{Current Sec}). Since each edge is at a constant potential, each edge carries a spin current $\pm\hbar v$ per momentum, as detailed above. Taking the thermal average of the total spin current in the low-temperature limit gives
\begin{equation}
    \begin{split}
        \expval{I^s_z} &= \int^\infty_{-\infty} dE\, \frac{\nu_0 }{2} \hbar v \left[f\left(E-\frac{eV}{2}\right) - f\left(E+\frac{eV}{2}\right)\right] \\
        &= \frac{e}{2\pi}V \,,
    \end{split} 
    \label{Thermal average}
\end{equation}
where $f$ is the Fermi function and $\nu_0 = 1/(\pi \hbar v)$ is the edge density of states per length. In this setup with a transverse voltage, we define the spin Hall conductance as $G^s_H = \expval{I^s_z}/V$. Since each edge is at a constant potential (Fig.~\ref{Conductance setups}a), the spin Hall conductance is quantized, $G^s_H = e/(2\pi)$, even in the presence of spin-non-conserving perturbations. This quantization can be traced back to the fact that the  spin current operator is determined by the local electron density, which does not change upon intra-edge backscattering at equilibrium. 

While the spin Hall conductance is robust against intra-edge backscattering, perturbations that couple modes on separate edges (inter-edge scattering) may result in reflections without a corresponding spin flip. The transfer of charge between the two edges changes the spin current, Eq.~(\ref{Spin current two edges}). Hence, such perturbations will lead to a decrease in the spin Hall conductance. To demonstrate this, we add an inter-edge scattering term to the two-edge Hamiltonian,
\begin{equation}
    \delta H = \int dx\, \gamma(x) \tilde{\Psi}^\dagger \tau_x \tilde{\Psi} \,.
    \label{Effective Hamiltonian two edges disorder}
\end{equation}
This perturbation conserves $S_z$ and therefore does not give rise to spin-torque. Nevertheless, since it does not conserve the number of particles on a given edge, it will lead to a non-quantized spin conductance.

As before, in order to define a scattering problem, we will assume that   $\gamma(x)$ is non-zero only in the interval $0 < x <x_0$. Since there are four edge modes in the two-edge system, we can promote  $r$ and $t$ in the scattering matrix $\mathbb{S}$ in Eq.~(\ref{Scattering matrix}) to  $2 \times 2$ matrices. In this case, $r_{ij}$ ($t_{ij}$) denote the amplitude of an incoming state from edge $j$ reflecting (transmitting) into an outgoing state on edge $i$. The nonzero components of $r$ and $t$ are
\begin{align}
    r_{12} &= r_{21} = -i\tanh{\eta_\gamma} \,, \label{Coupled edges reflections} \\
    t_{11} &= t_{22} = \sech{\eta_\gamma} \,, \label{Coupled edges transmissions}
\end{align}
where $\eta_\gamma = \int_0^{x_0} \gamma(x) \,dx/(\hbar v)$. The other components, meanwhile, vanish due to the lack of a term coupling states of opposite spin. Noting that the reflected edge modes now carry an opposing spin current to the incident and transmitted modes, we find that
\begin{align}
     &I^s_z(0) = \frac{\hbar v}{2}\left(1 - \abs{r_{12}}^2\right) = \frac{\hbar v}{2}\left(1 - \tanh^2{\eta_\gamma}\right) \,, \label{Coupled edges left current} \\
     &I^s_z(x_0) = \frac{\hbar v}{2}\abs{t_{11}}^2 = \frac{\hbar v}{2}\left(1 - \tanh^2{\eta_\gamma}\right) \,. \label{Coupled edges right current}
\end{align}
Hence, unlike intra-edge spin-flip perturbations, inter-edge tunneling without a spin flip conserves the spin current but results in a decrease of its value. As a result, in the spin-Hall setup, Fig.~\ref{Conductance setups}a, the spin Hall conductance $G^s_H$ is \textit{not} robust against inter-edge scattering. As was mentioned above, this result could be expected from the fact that the spin current couples to the difference of the density operators  between the two edges, Eq.~(\ref{Spin current two edges}), and the inter-edge scattering does not conserve this difference.

When an edge is not at constant potential but has a potential drop $V$ along it (left-right bias), the spin current can have a jump in the presence of spin-flip perturbations, as is illustrated by  Eqs.~(\ref{Magnetic disorder left current})--(\ref{Magnetic disorder right current}). This jump can be thought of as resulting from a non-zero spin torque, Eq.~(\ref{Discontinuity}), in the non-equilibrium setup. Due to this jump, one must define separate spin conductances, which we call incident ($G^s_I = \langle I^s_z(0) \rangle / V$) and transmitted ($G^s_T = \langle I^s_z(x_0) \rangle / V$), for current flowing on either side of the disordered region (see Fig.~\ref{Conductance setups}b). Even without inter-edge scattering, these conductances are not quantized in the presence of magnetic disorder (unlike $G^s_H$); their sum, however, is robust since $G^s_I + G^s_T =  G^s_H$, see Eq.~(\ref{No edge coupling relation 2}) below. 
Finally, we note that when there is a voltage drop $V$ across both edges and no top-bottom voltage, we expect no net spin current (see Fig.~\ref{Conductance setups}c). This case is the conventional two-terminal charge transport setup, and we define the corresponding two-terminal charge conductance $G^c_{2T}$ as a reference.

The above results that were derived for the simple models of Eq.~(\ref{Effective Hamiltonian disorder}) and Eq.~(\ref{Effective Hamiltonian two edges disorder}) illustrate the generic behavior of the spin conductances. We corroborate the findings by our numerical transport simulation discussed in Sec.~\ref{Results Sec}, where we simulate magnetic disorder as well as a quantum point contact (QPC) system to couple the edges (see Fig.~\ref{QPC system}). Before that, we introduce spin conductances defined in a four-terminal setup, Sec.~\ref{Current Sec}.

\section{\label{Current Sec} Multiterminal transport}

\begin{figure*}
    \centering
    \includegraphics[width=.98\textwidth]{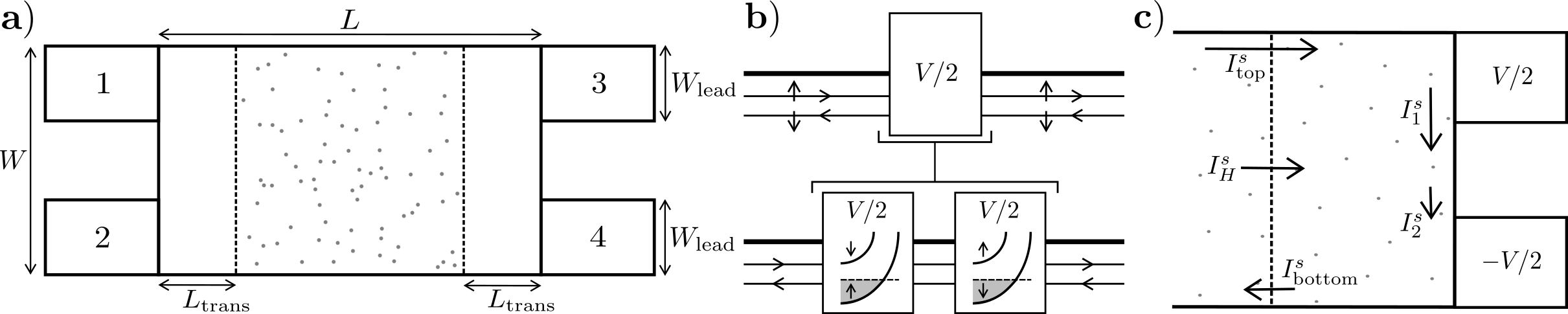}
    \caption{\textbf{a)} Schematic depiction of a four-terminal TI device of length $L$, width $W$, and lead width $W_{\text{lead}}$. The dotted region denotes disorder localized between two clean transition regions of length $L_{\text{trans}}$, where $L_{\text{trans}} = 0$ indicates a fully disordered sample. We evaluate the spin conductances from the spin currents entering the terminals, see Eq.~(\ref{Multiterminal spin conductance}). \textbf{b)} In order to measure the spin current entering each terminal, we consider the terminals to be composed of two closely spaced ferromagnetic leads with magnetization axes parallel and anti-parallel to the quantized axis of the TI, see Eq.~(\ref{Multiterminal spin current}). \textbf{c)} Depiction of spin currents in the spin Hall setup, Fig.~\ref{Conductance setups}a, with spin-non-conserving disorder between leads 3 and 4. The spin Hall current $I^s_H$ passing through a cross section of the TI sample is conserved and equal to the sum of the currents along the top and bottom edges, $I^s_H = I^s_{\text{top}} + I^s_{\text{bottom}}$. However, spin-non-conserving disorder and a non-equilibrium distribution lead to a spin torque on the edge connecting terminals 3 and 4, see also Fig.~\ref{Conductance setups}b. Due to the spin torque, an additional current $\delta I^s_H = I^s_1 - I^s_2$ is generated and flows to leads 3 and 4, resulting in a total spin current $I^s_H + \delta I^s_H$ entering the terminals and a non-quantized $G_H^s$.}
    \label{Four-terminal device}
\end{figure*}

We now move from the two-terminal case to a multiterminal system. While a two-terminal TI system requires the use of a proximitizing ferromagnetic heterostructure to drive a net spin current~\cite{gotte2016pure}, a spin Hall current can be driven purely electrically in a multiterminal setup. In this section we therefore give the relevant expressions for the currents and conductances necessary to study multiterminal charge and spin transport.

Consider a general $n$-terminal system with metallic leads attached. The full scattering matrix $\mathbb{S}$ of such a system relates the coefficients of the incoming modes $\Psi_{in}$ to the outgoing modes $\Psi_{out}$ by $\Psi_{out}=\mathbb{S}\Psi_{in}$. In particular, the $ij$-th block $\mathbb{S}_{ij}$ is the scattering matrix for modes scattering from terminal $j$ to $i$. Furthermore, in the case that the leads share a spin-rotational symmetry along a given axis, we may choose a new eigenbasis which conserves this symmetry. In this basis, the scattering matrix takes the form $\mathbb{S}_{i\sigma, j\sigma'}$, where the $\sigma$ indices denote the spins of the incoming and outgoing modes.

The Landauer-B\"uttiker formula provides the charge current passing through a lead in the low temperature limit in terms of the voltages applied to the leads and the transmission coefficients $T_{ij}$ (from terminal $i$ to $j$):
\begin{equation}
    I^c_i = \frac{e^2}{2\pi\hbar}\sum_{j \neq i}\left(T_{ji}V_j-T_{ij}V_i\right) \,.
    \label{Landauer-Buttiker formula}
\end{equation}
In the case of spin-rotational symmetric leads, Eq.~(\ref{Landauer-Buttiker formula}) may easily be generalized to give the spin-resolved current in a lead by considering each lead spin channel as a separate terminal:
\begin{equation}
    I^r_{i\sigma} = \frac{e}{2\pi\hbar}\sum_{j\sigma' \neq i\sigma}\left(T_{j\sigma', i\sigma}V_j-T_{i\sigma, j\sigma'}V_i\right) \,,
    \label{Spin-resolved current}
\end{equation}
where the spin-resolved current $I^r_{i\sigma}$ is the outgoing current in lead $i$ due to electrons of spin $\sigma$. The charge and spin currents in each lead can then be related to these spin-resolved currents by
\begin{align}
    I^c_i &= e\left(I^r_{i\uparrow} + I^r_{i\downarrow}\right) \label{Multiterminal charge current} \,, \\
    I^s_i &= \frac{\hbar}{2}\left(I^r_{i\uparrow} - I^r_{i\downarrow}\right) \,. \label{Multiterminal spin current}
\end{align}
The above equations also suggest that spin current can be measured by using two ferromagnetic terminals fully polarized along the $z$ and $-z$ axes. The net current into each terminal will be effectively spin resolved and their difference gives the net spin current. In Fig.~\ref{Four-terminal device}b, we envision using this technique to measure the spin current into each terminal~\footnote{Alternatively, one can use the inverse spin Hall effect, demonstrated in Ref.~\cite{brune2012spin}.}.

\begingroup
\setlength{\tabcolsep}{4pt} 
\renewcommand{\arraystretch}{1.6} 
\begin{table}
    \centering
    \begin{tabular}{|c|l|}
        \hline
        2-Terminal & $G^c_{2T} = G^c_{31} + G^c_{32} + G^c_{41} + G^c_{42}$ \\ \hline
        Incident & $G^s_I = -\frac{1}{2}\left(G^s_{11} - G^s_{12} + G^s_{21} - G^s_{22}\right)$ \\ \hline
        Transmitted & $G^s_T = \frac{1}{2}\left(G^s_{31} - G^s_{32} + G^s_{41} - G^s_{42}\right)$ \\ \hline
        \multirow{2}{*}{Hall} & $G^s_H = \frac{1}{2}\left( G^s_{31} - G^s_{32} + G^s_{33} - G^s_{34} \right.$ \\
        & \hfill $\left. + G^s_{41} - G^s_{42} + G^s_{43} - G^s_{44} \right)$ \\ \hline
        Diagonal Hall~\cite{kane_quantum_2005} & $G^s_D = \frac{1}{2}\left(G^s_{31} - G^s_{34}\right)$ \\
         \hline
    \end{tabular}
    \caption{Conductance definitions for various voltage setups in a four-terminal device. The terminal indexing matches Fig.~\ref{Four-terminal device}; see Eqs.~(\ref{Multiterminal charge conductance})--(\ref{Multiterminal spin conductance}) for the matrix elements. The additional negative sign for the incident conductance ensures that positive current is defined to move to the right. Fig.~\ref{Conductance setups} depicts the biasing setups (except for $G_D^s$).}
    \label{Conductance definitions table}
\end{table}
\endgroup

In the scattering formalism, the conductance $\mathbf{G}$ of an $n$-terminal system is the $n \times n$ matrix relating the currents in the leads to the applied voltages. Assuming the leads share the same spin-rotational symmetry as the TI in the pristine limit, we define the $2n \times n$ spin-resolved conductance matrix $\mathbf{G}^r$ by the spin-resolved current response $I^r_{i\sigma}$ to a small voltage $V_j$ (setting all other voltages to zero): $G^r_{i\sigma,j} = I^r_{i\sigma} / V_j$. From this we then define the $n \times n$ charge and spin conductance matrices $\mathbf{G}^{c/s}$ by
\begin{align}
    G^c_{i, j} &= e\left(G^r_{i\uparrow,j} + G^r_{i\downarrow,j}\right) = I^c_i / V_{j} \,, \label{Multiterminal charge conductance} \\
    G^s_{i, j} &= \frac{\hbar}{2}\left(G^r_{i\uparrow,j} - G^r_{i\downarrow,j}\right) = I^s_i / V_{j} \,. \label{Multiterminal spin conductance}
\end{align}
By inverting the conductance matrices, one could also quantify the inverse Hall effect and the inverse spin Hall effect, where a voltage is generated by a charge or spin current, respectively.

While the conductance matrices in Eqs.~(\ref{Multiterminal charge conductance})--(\ref{Multiterminal spin conductance}) provide the current response resulting from any voltage configuration, it is more illuminating to define conductance values for specific voltage setups such as those depicted in Fig.~\ref{Conductance setups}. In Table~\ref{Conductance definitions table} we define several such conductance values for the four-terminal device depicted in Fig.~\ref{Four-terminal device}a: the standard two-terminal charge conductance $G^c_{2T}$ due a horizontal potential bias, the incident and transmitted spin conductances $G^s_{I/T}$ due to a vertical bias on a single side, the spin Hall conductance $G^s_H$ due to a vertical bias on both sides, and the diagonal spin Hall conductance $G^s_D$ due to a diagonal bias (this was considered in Ref.~\cite{kane_quantum_2005}). We note that in the case of $G^s_D$ there is a potential drop on every edge. This leads to $G^s_D$  being  less robustly quantized than $G_H^s$, see Sec.~\ref{Magnetic disorder sec}. 

It is important to recognize that the spin conductances defined in Table~\ref{Conductance definitions table} are defined with regards to the spin currents passing through the leads. In a multiterminal system with spin-non-conserving disorder this is \textit{not} the same as spin currents passing through a cross section of the TI sample. In Fig.~\ref{Four-terminal device}c we demonstrate this difference in the case of the spin Hall current and conductance. The net spin current into leads 3 and 4 on the right has two components: the spin Hall current from the left leads, $I^s_H$, and the extra spin current between leads 3 and 4, $\delta I^s_H$, \textit{generated} by the spin torque from spin-non-conserving disorder, see  Eq.~(\ref{Discontinuity}). In terms of these, the spin Hall conductance is $G^s_H = \left(I^s_H + \delta I^s_H \right)/V$. In general, $G^s_H$ is not  equal to the conductance corresponding to just the spin Hall current passing through the sample, $G^s_{H'} = I^s_H/V$, especially when the connection between leads 3 and 4 is disordered (see Sec.~\ref{Magnetic disorder sec}). Importantly, only $G^s_{H'}$ is quantized as predicted in Sec.~\ref{Effective Hamiltonian Sec} when the entire sample is disordered; $G^s_H$ is only quantized when the connection between leads 3 and 4 has no spin-symmetry breaking disorder~\footnote{In the spin Hall setup with no inter-edge scattering, the total spin current in leads 3 and 4 is directly related to the \textit{charge} current $I^c_{34} = (e^2/h)\, T_{34}V$ in the connection between them: $I^s_3 + I^s_4 = (\hbar/e)\,I^c_{34}$. Hence, defining the conductance of the right edge $G^c_{\text{right}} = I^c_{34}/V$ where $V$ is the voltage difference between leads 3 and 4, we have $G^s_H = (\hbar/e)\,G^c_{\text{right}} \leq e/(2\pi) = G^s_{H'}$.}. This picture is confirmed by our numerical study where we compare clean and disordered connection between leads 3 and 4, see Fig.~\ref{Magnetic disorder plots}.

Using the definitions provided by Table~\ref{Conductance definitions table}, we can derive several relations between the four-terminal conductances. In particular, we consider two special cases which will be relevant to the results in Secs.~\ref{Scalar disorder sec} and \ref{Magnetic disorder sec}. When the disorder does not break the spin-rotational symmetry of the TI, transmission between opposite spins is impossible: $T_{i\sigma,j\sigma'} \propto \delta_{\sigma\sigma'}$. This restriction results in the following relations between the conductances,
\begin{align}
    G^s_H &= \frac{\hbar}{2e}G^c_{2T} \,, \label{No spin flip relation 1} \\
    G^s_I &= G^s_T = \frac{1}{2}G^s_H \,. \label{No spin flip relation 2}
\end{align}
The relations in Eqs.~(\ref{No spin flip relation 1})--(\ref{No spin flip relation 2}) are valid so long as every conducting state is spin-polarized and the spin-rotational symmetry remains unbroken. Meanwhile, if there is no inter-edge scattering then only spin-preserving transmission and spin-flipping reflections are allowed: $T_{i\sigma,j\sigma'} \propto \abs{\delta_{ij} - \delta_{\sigma\sigma'}}$. The resulting conductance relations are,
\begin{align}
    G^s_T &= \frac{\hbar}{4e} G^c_{2T} \,, \label{No edge coupling relation 1} \\
    G^s_H &= G^s_I + G^s_T \,. \label{No edge coupling relation 2}
\end{align}
Unlike Eqs.~(\ref{No spin flip relation 1})--(\ref{No spin flip relation 2}), the relations in Eqs.~(\ref{No edge coupling relation 1})--(\ref{No edge coupling relation 2}) rely on the localization of the edge states and are not true in the presence of conducting bulk states.

\section{\label{Results Sec}Numerical studies of disordered multiterminal systems}

\begin{figure}
    \centering
    \includegraphics[width=.5\textwidth]{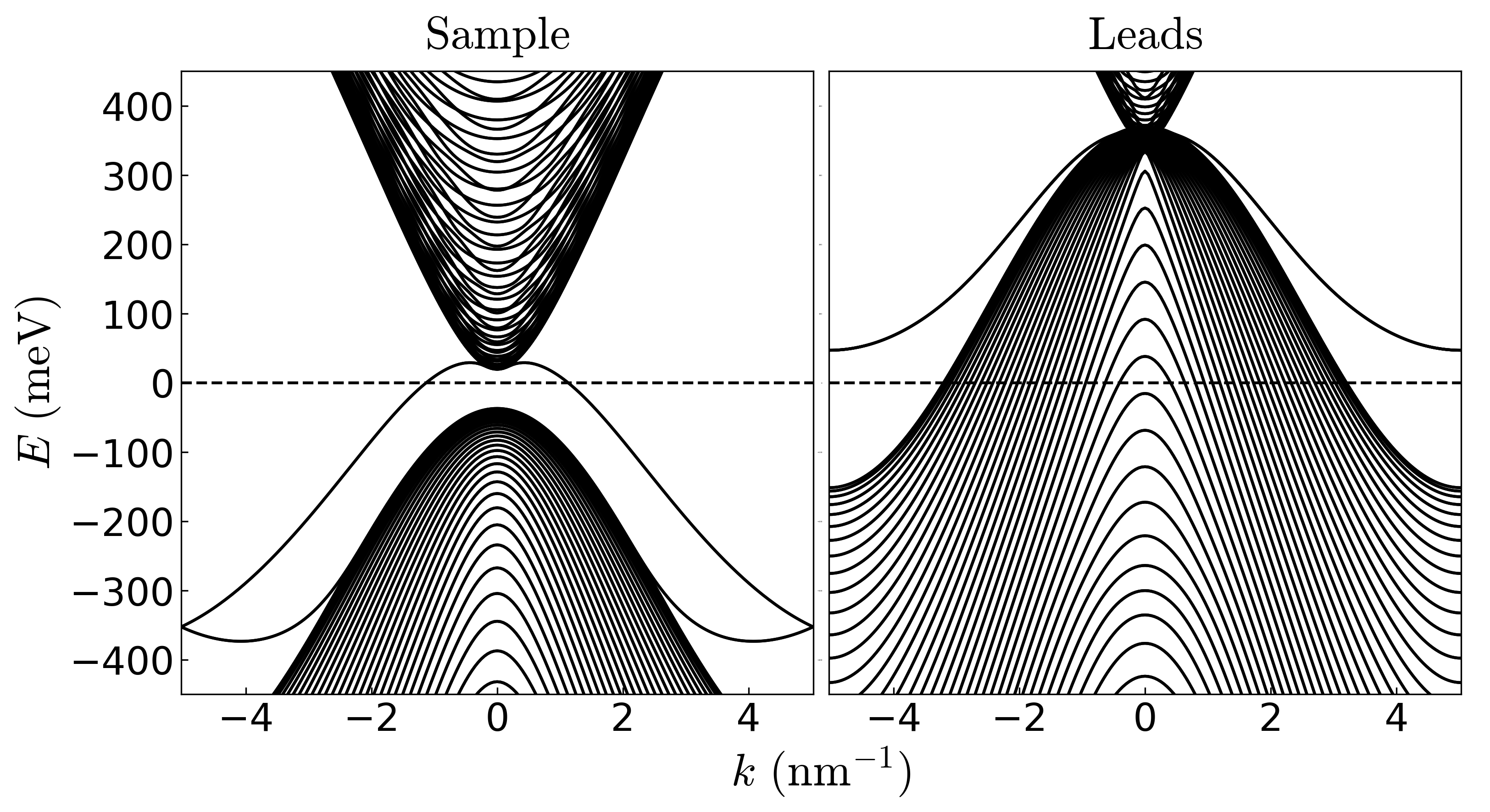}
    \caption{Straight-edge terminated WTe$_2$ band structure in a pristine sample (left) and leads (right), corresponding to the TI and metallic phases, respectively. The dashed line shows the chemical potential; the lead bands are shifted by $400$ meV relative to the sample.}
    \label{Band structure}
\end{figure}

To numerically study the transport properties of WTe$_2$, we utilized the Kwant package~\cite{Groth2014KwantAS} for Python to implement the tight-binding model introduced in Ref.~\cite{Lau2019}. Four-terminal systems were created to study the conductances in Table~\ref{Conductance definitions table}. Each system is comprised of a sample in the topological phase with four  leads of width $W_{\text{lead}} = 12$ nm attached at the corners, as depicted in Fig.~\ref{Four-terminal device}. We model the leads with the same WTe$_2$ tight-binding model as the sample, except with spin-orbit coupling set to zero. The Fermi level of the leads is placed within the valence band ($\mu = -400$ meV) to allow for an abundance of conducting bulk modes; the sample Fermi level, meanwhile, is placed near the center of the 56 meV wide bulk gap ($E=0$ in Fig.~\ref{Band structure}) to ensure only edge modes are relevant in the pristine, zero-temperature limit. All plots shown utilize a horizontal straight-edge termination~\footnote{We use the nomenclature from Ref.~\cite{Lau2019}. Depending on sample length $L$, the vertical edges are either W or Te-terminated which both have a buried Dirac point.} that has a Dirac point buried within the valence band (see Fig.~\ref{Band structure}); however, we find similar results for the zigzag termination which has a Dirac point in the bulk gap. We then use Kwant to construct the scattering matrix for the system, which is used with Eqs.~(\ref{Spin-resolved current})--(\ref{Multiterminal spin conductance}) to determine the charge and spin conductances in the zero-temperature limit~\footnote{Our code to reproduce the figures is available at: \url{https://purr.purdue.edu/publications/3929/1}.}.

Unless otherwise stated, each plot represents the average of $N=300$ disordered samples, which we find to be enough to limit most fluctuations (see Appendix~\ref{Convergence sec}). We also attach the standard error bars for each plot (i.e. $\pm\sigma_G/\sqrt{N}$). For each plot we measure the conductances in terms of the charge and spin conductance quanta, $G^c_0 = e^2/h$ and $G^s_0 = e/(4\pi)$, respectively.

In the pristine limit we find the standard~\cite{kane_quantum_2005} quantized values for the two-terminal charge conductance ($G^c_{2T} = 2e^2/h$) and spin Hall conductance ($G^s_H = e/(2\pi)$). We also find that $G^s_I = G^s_T = e/(4\pi)$ and $G^s_D = e/(4\pi)$~\cite{kane_quantum_2005} in the pristine limit. In the following subsections we discuss the effects of on-site scalar and magnetic disorder on these results, in addition to disorder in the spin-orbit coupling parameters. We also study inter-edge scattering using a QPC system and calculate the characteristic spin decay length in the presence of magnetic disorder.

\subsection{\label{Scalar disorder sec}$S_z$ conserving disorder}

\begin{figure}
    \centering
    \includegraphics[width=.48\textwidth]{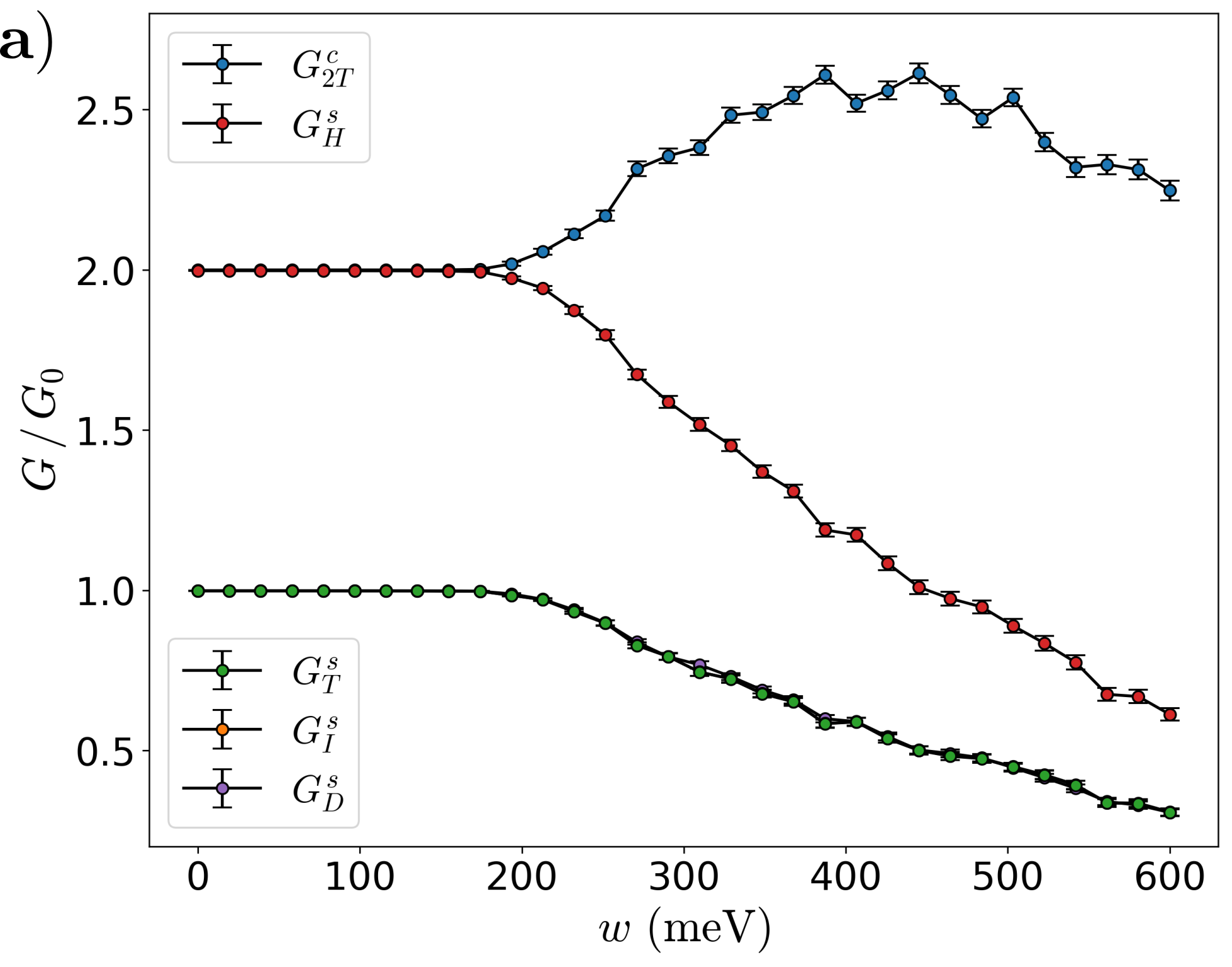} \\
    \includegraphics[width=.48\textwidth]{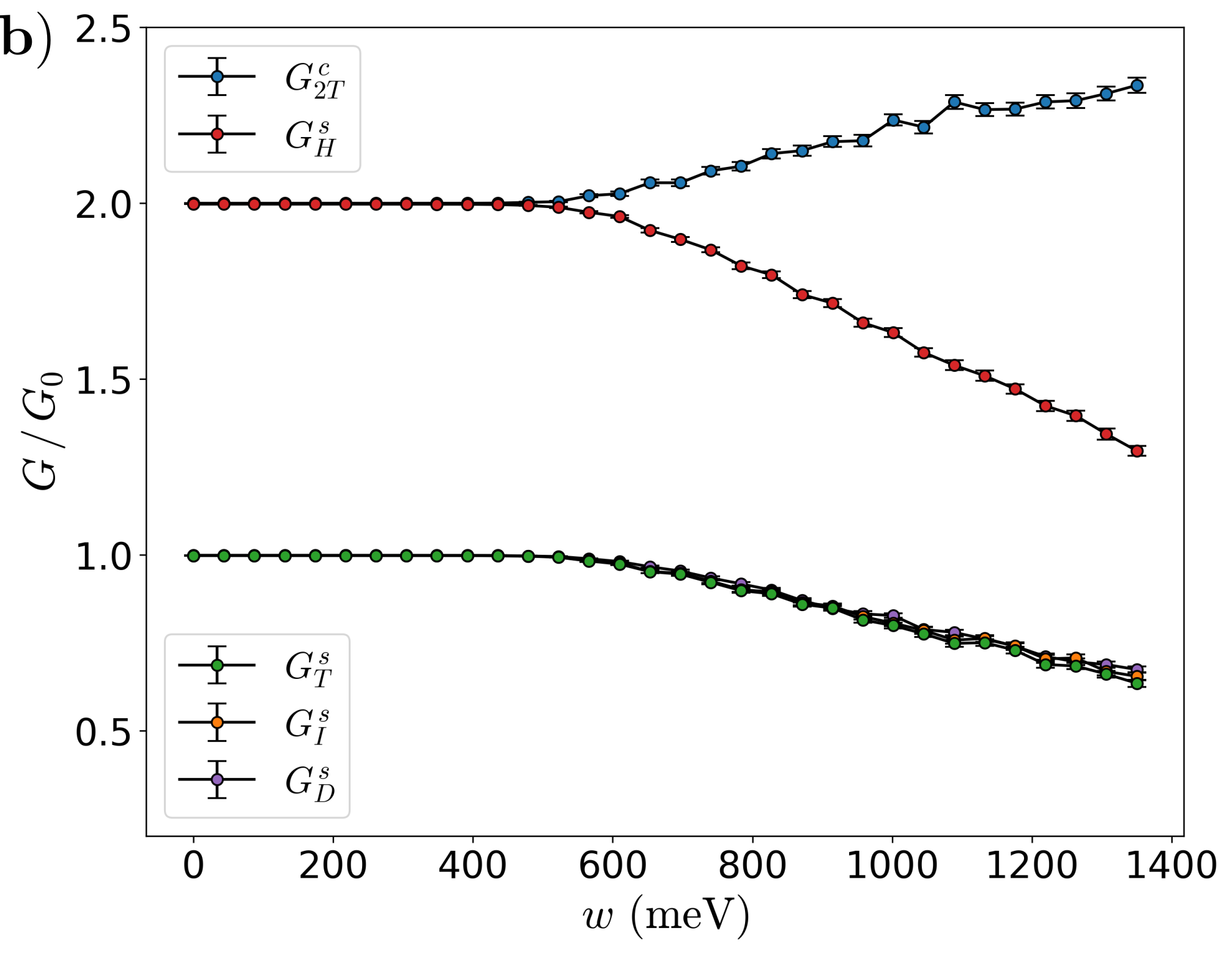}
    \caption{Conductances versus disorder strength $w$ for spin-conserving TR-symmetric perturbations. The system dimensions are $L=20$ nm, $W=30$ nm, $L_{\text{trans}}=0$, and $W_{\text{lead}}=12$ nm (see Fig.~\ref{Four-terminal device}). Conductances are measured in units of the charge and spin conductance quanta. Note that the lowest 3 curves, $G^s_T$, $G^s_I$, and $G^s_D$,  are overlapping over the full range of $w$ in both figures. \textbf{a)} On-site scalar disorder. \textbf{b)} Spin-conserving SOC disorder.}
    \label{Scalar disorder plots}
\end{figure}

Due to the spin-momentum locking of the edge states in a 2D TI, it is expected that any perturbation which neither breaks the spin-symmetry nor couples the edges will not affect current propagation, as long as the perturbation strength is smaller than the gap to bulk excitations. Previous studies~\cite{Lau2019, PhysRevLett.102.136806} have demonstrated this in the context of scalar disorder and charge conductance. Here, we demonstrate that weak spin-symmetric disorder does not affect the charge and spin conductance values of our four-terminal system. We study the effects of both on-site scalar disorder as well as disorder in the SOC strength.

In Fig.~\ref{Scalar disorder plots}a we add a spatially-dependent on-site potential $u(\mathbf{x})$ drawn from a Gaussian of mean 0 and standard deviation $w$; we then plot the dependence of the conductances defined in Table~\ref{Conductance definitions table} on $w$. For small enough values of $w$ ($< 200$ meV), we find that the charge and spin conductances remain quantized at their expected values. This is due to the fact that scalar on-site disorder does not break the TR and spin-rotational symmetries of the TI, nor does it couple the two edges; the transmission amplitudes thus remain unaffected when the disorder is weak. At larger $w$, however, we see a decrease in the spin conductances and an increase in the charge conductance. The increasing charge conductance is attributable to the onset of bulk conduction within the disordered sample, whose size is smaller than the Anderson localization length. For weak disorder, the Fermi level of the sample remains within the bulk gap, ensuring that only the spin-momentum locked edge states effect the low-temperature conductances. Stronger disorder, meanwhile, can shift the bands sufficiently so that they cross the Fermi level, leading to bulk conduction. 

The effect of disorder in the SOC strength is similar to spin-symmetric on-site disorder. In Fig.~\ref{Scalar disorder plots}b we multiply the SOC strength by a spatially dependent factor $\lambda(\mathbf{x})$ drawn from a Gaussian of mean 1 and standard deviation $\delta\lambda$; we then plot the conductances versus $w = \lambda_{SOC}\delta\lambda$, where $\lambda_{SOC} = 225$ meV is the sum of the SOC parameter magnitudes in the WTe$_2$ tight-binding model~\cite{Lau2019} (see Appendix~\ref{TB model sec} for details on the WTe$_2$ tight-binding model). Importantly, this ``isotropic''  modification of the SOC strength does not change the spin quantization axis; this is unlike with anisotropic SOC disorder, see Sec.~\ref{Time-reversal symmetric  disorder sec} below. Just as with spin-symmetric on-site disorder, the conductances are robust against weak spin-symmetric SOC disorder; however, this regime appears to be smaller for SOC disorder, with the conductances deviating from their quantized values for $w > 60$ meV.

\begin{figure}
    \centering
    \includegraphics[width=.48\textwidth]{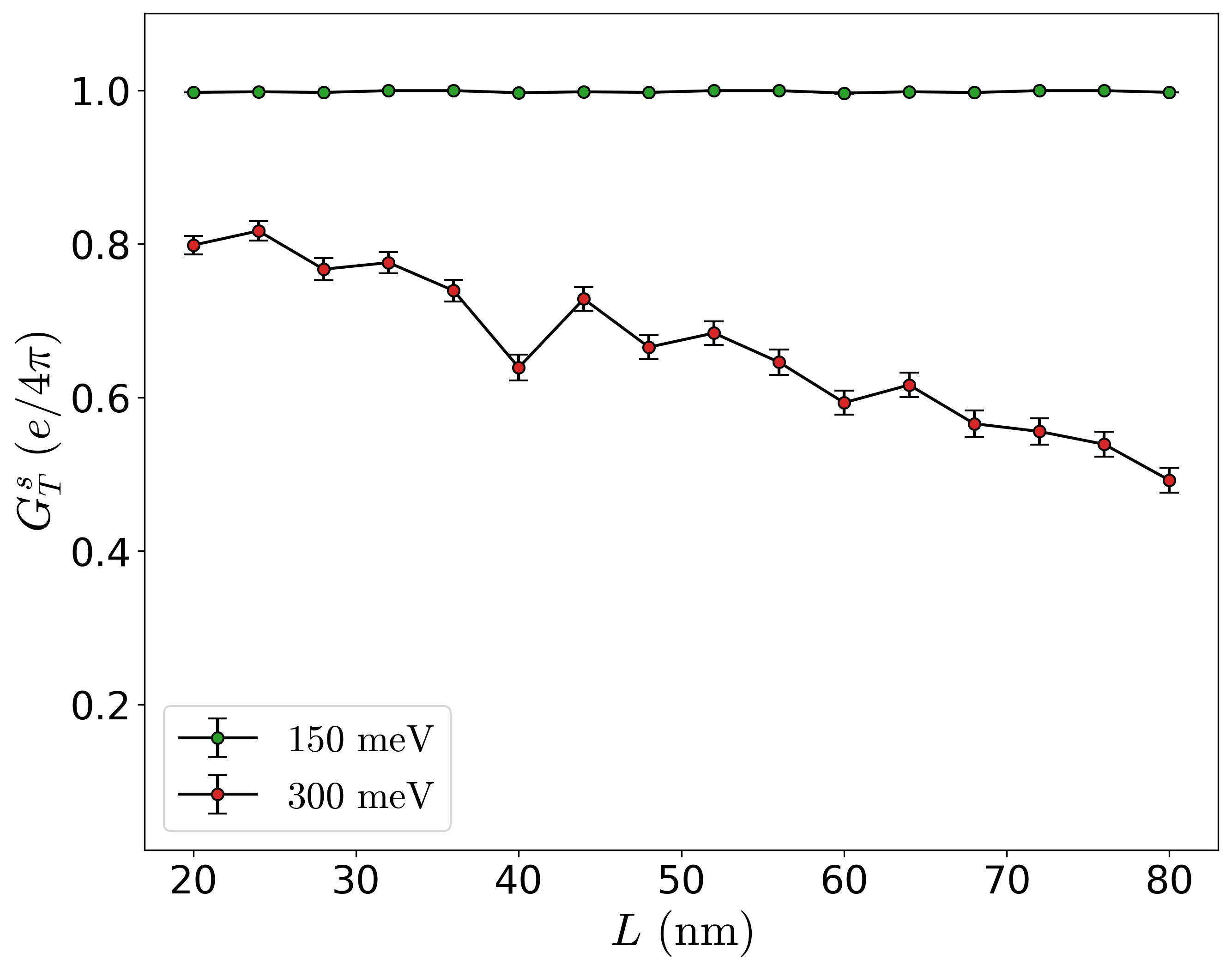}
    \caption{Transmitted spin conductance $G^s_T$ with on-site scalar disorder of width $w = 150$ meV or $w = 300$ meV versus sample length $L$. The other sample dimensions are $W=30$ nm, $L_{\text{trans}}=3$ nm, and $W_{\text{lead}}=12$ nm (see Fig.~\ref{Four-terminal device}).}
    \label{Scalar spin decay}
\end{figure}

The conductances are remarkably robust against weak spin-symmetric disorder. In Fig.~\ref{Scalar spin decay} we plot the transmitted spin conductance $G_T^s$ versus sample length for $w = 150$ meV and $w = 300$~meV on-site scalar disorder. In the weak disorder regime, the conductance remains quantized  and does not appear to depend on the length up to $L=100$~nm (not shown). Weak length-dependence appears in the very strong disorder regime ($w > 200$ meV for on-site scalar disorder). These findings are to be contrasted with a diffusive conductor where the conductance is inversely proportional to the length.

\subsection{\label{Time-reversal symmetric  disorder sec} Time-reversal symmetric, $S_z$ non-conserving disorder}

\begin{figure}
    \centering
    \includegraphics[width=.48\textwidth]{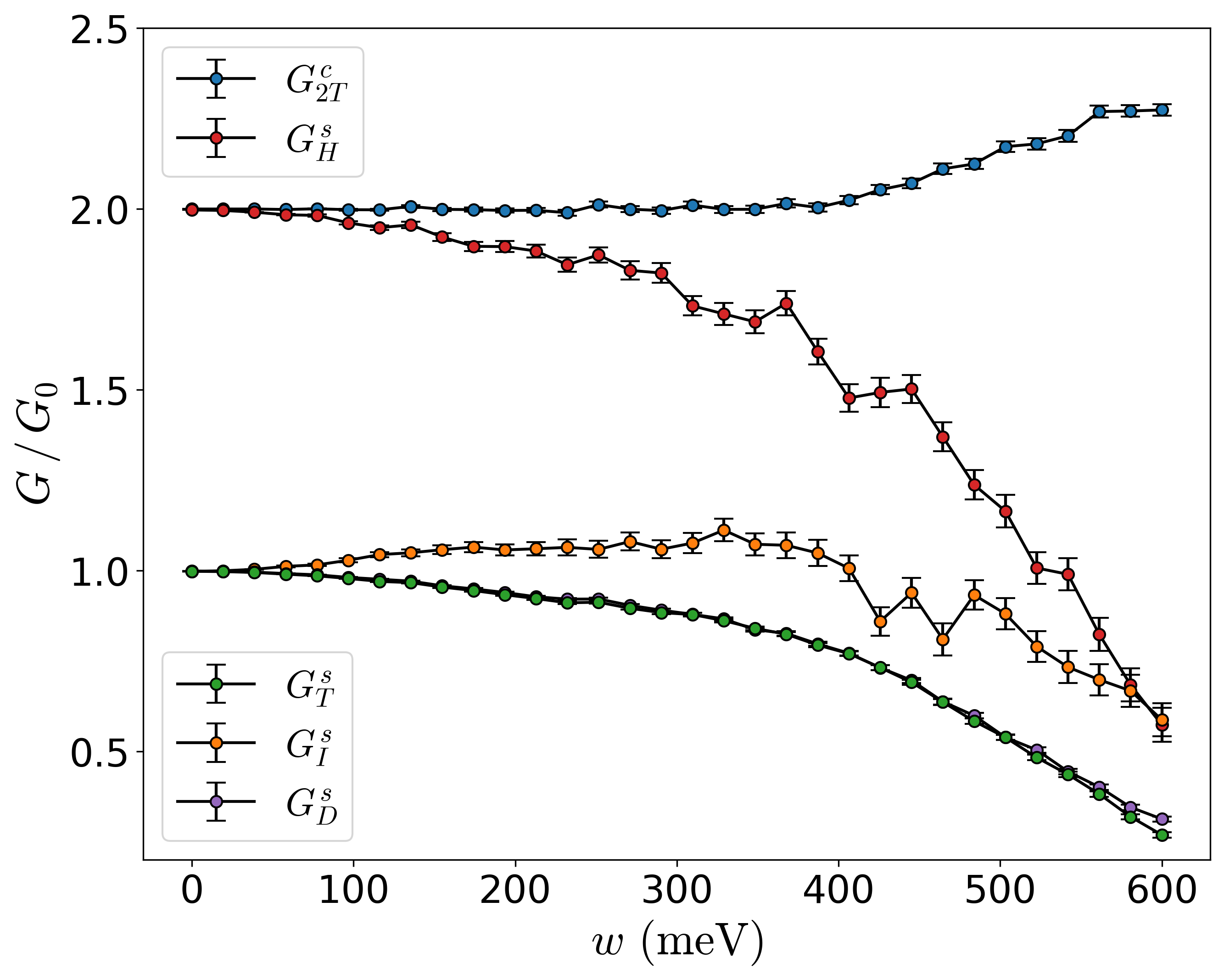}
    \caption{Conductances versus spin-non-conserving SOC disorder width $w$. The system dimensions are $L=20$ nm, $W=30$ nm, $L_{\text{trans}}=0$ nm, and $W_{\text{lead}}=12$ nm (see Fig.~\ref{Four-terminal device}). Each data point represents the average of 500 samples. Conductances are measured in units of the charge and spin conductance quanta. Note that the $G^s_T$ and $G^s_D$ curves are overlapping over almost the full range of $w$.}
    \label{TR symmetric plot}
\end{figure}

In Sec.~\ref{Scalar disorder sec} we saw that the charge and spin conductances remained quantized in the presence of weak on-site and SOC perturbations that do not break the spin-rotational symmetry of the TI. Here, we demonstrate that the conductances are \textit{not} protected against SOC perturbations that break the spin-rotational symmetry, even when TR symmetry remains intact. In particular, we implement a TR-symmetric, $S_z$ non-conserving disorder term by adding a spatially-dependent $i\lambda'_{0,x}(\mathbf{x})\sigma_x$ term to the $\lambda'_0$ hopping amplitude, where 
$\lambda'_{0,x}(\mathbf{x})$ is drawn from a Gaussian of mean 0 and standard deviation $w$ (see Appendix~\ref{TB model sec}). We demonstrate the effects of this term on the conductances in Fig.~\ref{TR symmetric plot}. As expected, SOC disorder that breaks $S_z$ conservation (Fig.~\ref{TR symmetric plot}) will lead to a stronger suppression of edge spin conductances as opposed to $S_z$ conserving SOC disorder (Fig.~\ref{Scalar disorder plots}b).

For disorder terms weaker than $w < 300$ meV, the conductances slowly deviate from their quantized values. This result suggests that TR symmetry alone is not enough to ensure quantization of the spin conductances when disorder is added to the SOC hopping amplitudes; rather, it is the combination of TR symmetry and spin-rotational symmetry that leads to this quantization. Of course, this distinction is not relevant when one only considers on-site disorder terms, as in that case spin-rotational symmetry is implied by TR symmetry. At larger $w$ we see a qualitatively different dependence of conductance on disorder strength, corresponding to the onset of bulk conduction in the disordered sample. While the conductances do not remain quantized in the presence of TR-symmetric, spin-non-conserving disorder, their deviations from their quantized values appears to be much weaker than for disorder that breaks TR symmetry, see Fig.~\ref{Magnetic disorder plots}b in Sec.~\ref{Magnetic disorder sec}.

\subsection{\label{Magnetic disorder sec}Magnetic disorder breaking time-reversal symmetry and $S_z$ conservation}

\begin{figure}
    \centering
    \includegraphics[width=.48\textwidth]{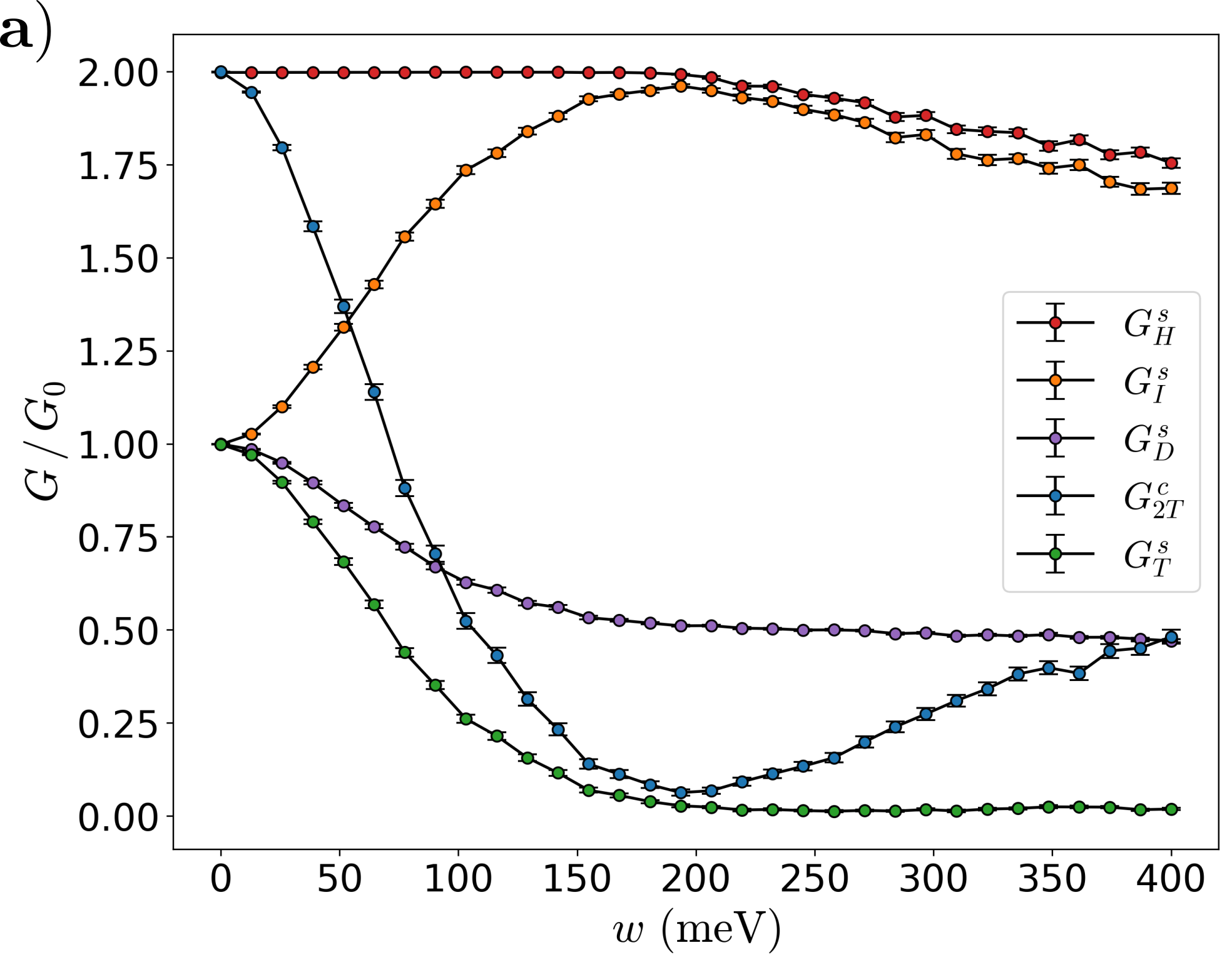} \\
    \includegraphics[width=.48\textwidth]{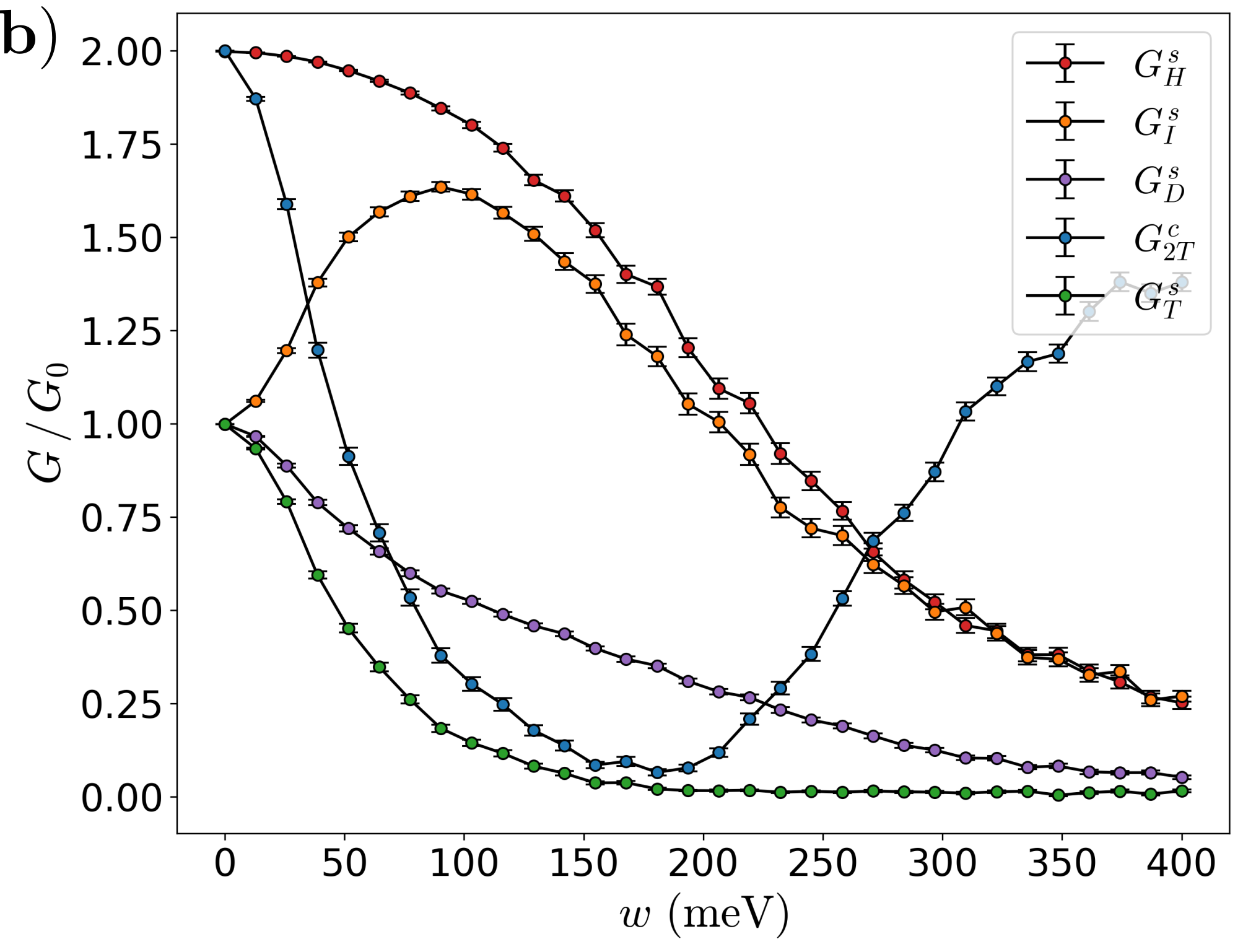}
    \caption{Conductances versus on-site magnetic disorder width $w$. For both plots $L=20$ nm, $W=30$ nm, and $W_{\text{lead}}=12$ nm (see Fig.~\ref{Four-terminal device}). Conductances are measured in units of the charge and spin conductance quanta. \textbf{a)} A $L_{\text{trans}}=2.5$ nm wide clean transition region is added to the ends of the TI to ensure no disorder at the lead-TI interfaces. In this case the spin current entering the terminals is approximately conserved and $G^s_H$ stays quantized up to large $w\lesssim 200$ meV. \textbf{b)} No such transition region is added, $L_{\text{trans}}=0$. In this case there is a spin torque that prevents the quantization of $G^s_H$, see Fig.~\ref{Four-terminal device}c.}
    \label{Magnetic disorder plots}
\end{figure}

Unlike spin-symmetric on-site disorder and SOC disorder, unaligned magnetic disorder breaks both the TR symmetry and the spin-rotational symmetry of the TI, leading to a large deviation of the conductance from the pristine-limit quantization even before the onset of bulk conduction. To demonstrate this, we add a $m(\mathbf{x})\sigma_x$ on-site disorder term, where $m(\mathbf{x})$ is once again drawn from a Gaussian of mean 0 and standard deviation $w$. We also show how the conductances defined in the leads depend drastically on whether or not there is disorder along the left and right edges.

In Fig.~\ref{Magnetic disorder plots}a we demonstrate the case of magnetic disorder localized such that there is no disorder between leads of the same side ($L_{\text{trans}}=2.5$ nm in Fig.~\ref{Four-terminal device}). We see that the spin Hall conductance $G^s_H$ maintains its quantized value until the onset of bulk conduction at about $w = 200$ meV, demonstrating the robustness predicted by Eq.~(\ref{Thermal average}). Meanwhile, the charge conductance $G^c_{2T}$ and transmitted spin conductance $G^s_T$ immediately begin to decrease with $w$ while the incident spin conductance $G^s_I$ increases. These deviations are in qualitative agreement with Eqs.~(\ref{Magnetic disorder left current})--(\ref{Magnetic disorder right current}) if we make the identification $\eta_m = w^2x_0r_0/(\hbar^2 v^2)$, where $x_0 = L - 2L_{\text{trans}}$ is the length of the disordered region and $r_0$ is the correlation length of the disorder. The conductances also obey the relations predicted by Eqs.~(\ref{No edge coupling relation 1})--(\ref{No edge coupling relation 2}). Similarly, the diagonal spin Hall conductance $G^s_D$ deviates from its quantized value at a much lower strength of disorder than $G^s_H$. We attribute this difference to the different biasing configurations: in measuring $G^s_D$, every edge has a voltage drop which allows for large spin torque contributions (see Sec.~\ref{Effective Hamiltonian Sec}). We also note that $G^s_D$ appears to decrease to half of its zero-disorder quantized value. This is due to the fact that in Table~\ref{Conductance definitions table} for very strong disorder $G^s_{31} \to 0$ but $G^s_{34} = -1$ due to the clean connection between leads 3 and 4. 

Meanwhile, in Fig.~\ref{Magnetic disorder plots}b, we demonstrate the case of a fully-disordered sample with magnetic disorder added along the edges connecting leads of the same side ($L_{\text{trans}}=0$ in Fig.~\ref{Four-terminal device}). We see that the removal of the clean connection results in a different dependence on the disorder strength. The relations given by Eqs.~(\ref{No edge coupling relation 1})--(\ref{No edge coupling relation 2}), which only relied on the lack of bulk conduction and edge-to-edge coupling, still hold for $w < 200$ meV. However, the spin Hall conductance $G^s_H$ is apparently no longer quantized, and the deviations of $G^s_I$ and $G^s_T$ no longer agree with what is predicted by Eqs.~(\ref{Magnetic disorder left current})--(\ref{Magnetic disorder right current}). As mentioned in Sec.~\ref{Current Sec}, this discrepancy is due to the fact that we define the conductances in the leads, not in the sample. We expect the spin Hall conductance corresponding to the current in the sample to remain quantized even when the sample is strongly disordered.

\begin{figure}
    \centering
    \includegraphics[width=.48\textwidth]{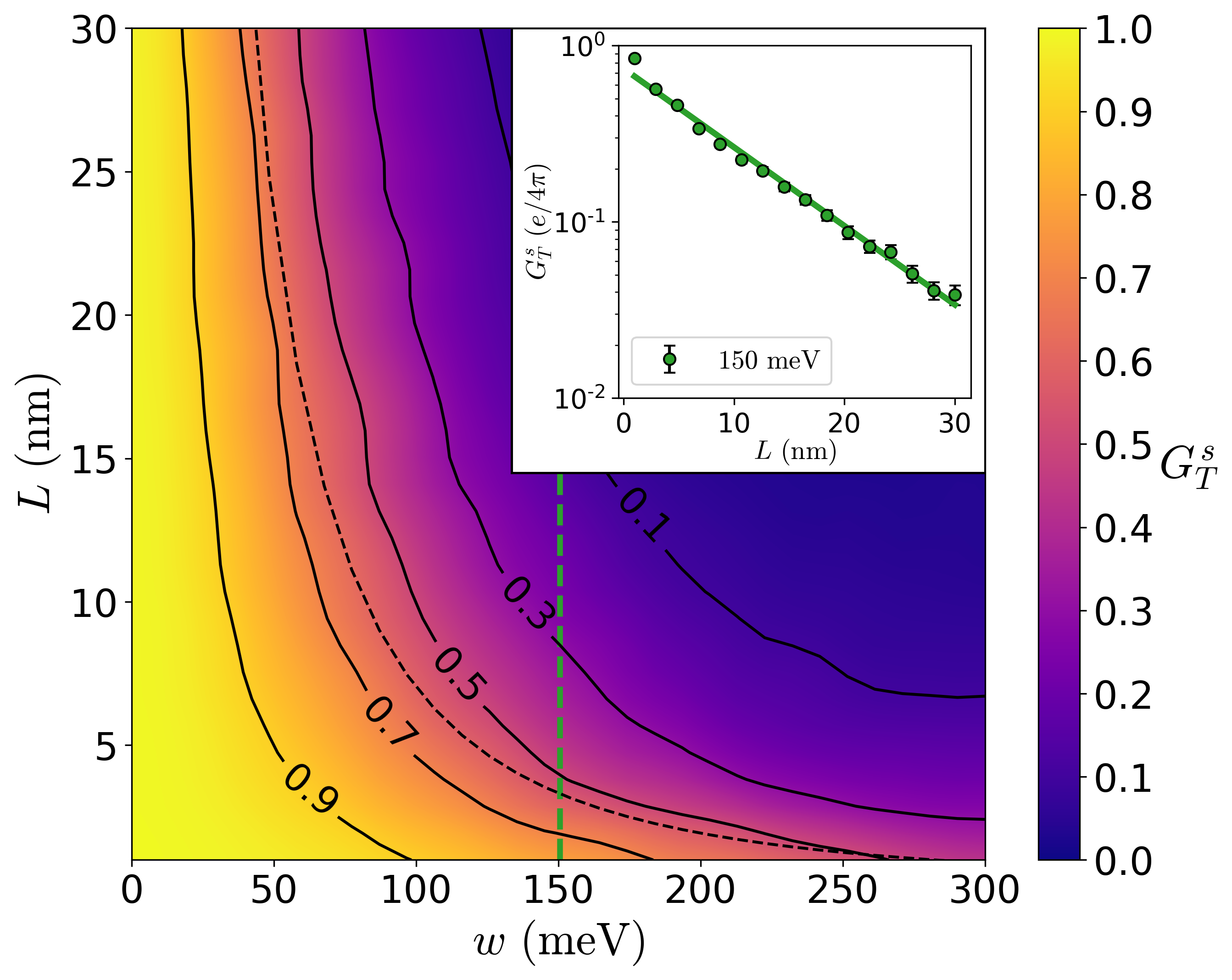}
    \caption{Transmitted spin conductance $G^s_T$ as a function of sample length $L$ and on-site magnetic disorder width $w$ for a 32 by 32 mesh grid. The other sample dimensions are $W=30$ nm, $L_{\text{trans}}=3$ nm, and $W_{\text{lead}}=12$ nm (see Fig.~\ref{Four-terminal device}a). The average value of $G^s_T$ over 50 samples was used to color each grid point; the plot was then smoothed using a Gaussian. The solid lines are contours of constant $G^s_T$ and roughly follow a $L \propto w^{-2}$ dependence (black dashed line), as is predicted by the relation $\ln G^s_T \propto -L w^2$. \textit{Inset:}~Logarithmic plot of $G^s_T$ averaged over 300 samples versus $L$ for a fixed $w = 150$ meV slice (green dashed line in the main figure). The slope of the best fit line (solid green) is $-1/(9.7\,\text{nm})$.}
    \label{Spin decay plots}
\end{figure}

In addition to studying how the disorder strength affects the conductances, we also study how the transmitted conductance $G^s_T$ varies with the sample length $L$. We plot the dependence of $G^s_T$ on the disorder strength and sample length, as well as a constant $w = 150$ meV slice, in Fig.~\ref{Spin decay plots}. We find that, for constant $w$, the transmitted spin conductance decays exponentially with the sample length, i.e. $G^s_T \propto e^{-L/l_0}$ where $l_0$ is a characteristic spin decay length. For $w = 150$ meV, our fit gives $l_0 \approx 9.7$ nm, see inset of Fig.~\ref{Spin decay plots}. This roughly agrees with an estimate of $l_0 = \hbar^2v^2/(w^2r_0) \approx 3.2$ nm if we use the average distance between neighboring lattice sites $r_0 \approx .2$ nm as a disorder correlation radius and $\hbar v \approx 120 \,\mathrm{meV} \cdot \mathrm{nm}$ estimated from Fig.~\ref{Band structure}.

\subsection{\label{QPC sec}Quantum point contact system}

\begin{figure}
    \centering
    \includegraphics[width=.48\textwidth]{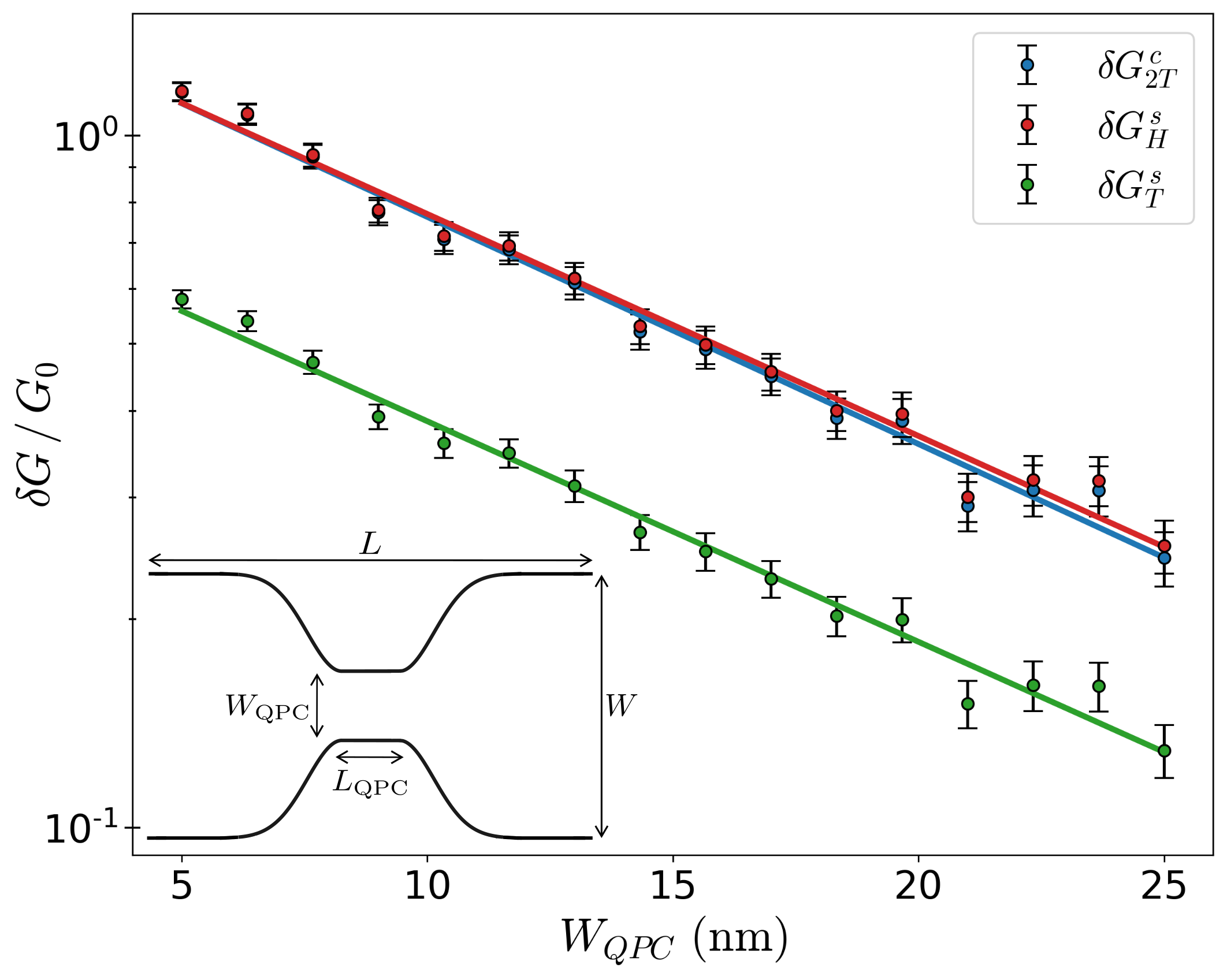}
    \caption{Conductance deviations $\delta G = G_0 - G$ of a four-terminal QPC system made of a $L=30$ nm by $W=30$ nm rectangular sample cut such that the width smoothly transitions to a narrowed region of length $L_{\text{QPC}} = 200$ nm and varying width $W_{\text{QPC}}$. A scalar disorder term with standard deviation $w = 300$ meV is then added to extend the effective decay length and increase edge-to-edge coupling. Conductance deviations are measured in units of the charge and spin conductance quanta. The slopes of the corresponding best fit lines are $-1/(13.2\,\text{nm})$ for $G^c_{2T}$, $-1/(13.6\,\text{nm})$ for $G^s_H$, and $-1/(13.5\,\text{nm})$ for $G^s_T$. \textit{Inset:} Diagram of the QPC device demonstrating the definitions of the various dimensions.}
    \label{QPC system}
\end{figure}

As mentioned in Sec.~\ref{Effective Hamiltonian Sec}, inter-edge tunneling through the bulk of the TI is another mechanism by which the conductances can deviate from their quantized values. For each conductance $G$ we define the deviation $\delta G$ from the quantized value $G(w=0)$ by $\delta G = G(w=0) - G$. In a QPC system of minimum width $W_{\text{QPC}}$, we expect $\delta G \propto e^{-W_{\text{QPC}}/W_0}$ for $ W_{\text{QPC}} \gg W_0$, where $W_0$ is the effective decay length of the edge modes (not to be confused with the characteristic spin decay length $l_0$ studied in Sec.~\ref{Magnetic disorder sec}). To test this relation, we create a four-terminal QPC system where a rectangular sample is smoothly transitioned into a narrowed region of width $W_{\text{QPC}}$ and length $L_{\text{QPC}}$ (see inset of Fig.~\ref{QPC system}). We then add a scalar disorder term to extend the effective decay length $W_0$.

In Fig.~\ref{QPC system} we plot the resulting conductance deviations against $W_{\text{QPC}}$ on a logarithmic scale, along with their linear fits. Using the inverse slopes of the best fit lines, we find that the decay lengths of each conductance component is roughly 13 nm. The various spin conductance deviations, including the incident and diagonal conductance deviations which we hide for clarity, have similar decay lengths. 
Physically, the decay length serves as an indicator of the edge state width in the QPC geometry. We note that each conductance component decays at the same rate as is expected from Eqs.~(\ref{No spin flip relation 1})--(\ref{No spin flip relation 2}), valid for a system with spin conservation~\footnote{For small $\delta G \lesssim 10^{-2}$, such as in the case of weaker disorder, we find a larger decay length for $\delta G^c_{2T}$ as compared to  $\delta G^s_{H}$. We attribute this ``violation'' of Eq.~(\ref{No spin flip relation 1}) to our inaccurate determination of the spin quantization axis, see Sec.~\ref{TB model sec}.}.

\section{Conclusions \label{Conclusions Sec}}

We studied  the effects of disorder on spin transport in 2D TIs and  established important estimates for the level of  disorder strength that starts to hinder spin transport. One of our main findings is that the spin current operator on the 2D TI edge is given by the local density, Eq.~(\ref{Spin current two edges}). For this reason, the spin Hall current generated by a transverse voltage is remarkably robust to even spin-non-conserving perturbations, see Eq.~(\ref{Thermal average}), as long as the two edges of the 2D TI are not coupled. However, measuring the spin Hall current in a 4-terminal geometry is difficult due to additional spin currents that flow between the terminals at different potentials, see Fig.~\ref{Four-terminal device}c. These spin currents are not in general conserved and hinder the measurement of a quantized spin Hall conductance. These findings are confirmed by our numerical simulations, e.g. Fig.~\ref{Magnetic disorder plots}. Overall, we find that spin conductance is most sensitive to spin-non-conserving disorder such as random spin-orbit coupling (Fig.~\ref{TR symmetric plot}) or magnetic impurities (Figs.~\ref{Magnetic disorder plots}--\ref{Spin decay plots}). In the former time-reversal symmetric case, the spin Hall conductance is nevertheless nearly quantized even with relatively large disorder strength of the order of the bulk band gap. 

In WTe$_2$, recent measurements of the spin quantization axis indicate that spin-orbit disorder is relatively weak. The canting of the edge state spin has been measured in experiments~\cite{tan2020determination,2020arXiv201009986Z} in agreement with theoretical models~\cite{2020arXiv200705626G,PhysRevB.102.161402,2019PhRvB..99l1105O,Lau2019, nandy2021lowenergy}. These findings indicate that the spin quantization axis, although canted, does not vary strongly in position or momentum space. This gives hope that the spin of the edge carriers can be conserved over long distances. 

We focused on low-temperatures at which scattering is dominated by elastic processes. At the same time, we found that time-reversal symmetric disorder  has a weak effect on spin transport, see Secs.~\ref{Scalar disorder sec}--\ref{Time-reversal symmetric  disorder sec}. Therefore, at higher temperatures, inelastic scattering is expected to become the dominant scattering mechanism, leading to temperature-dependent corrections to the spin conductances. Finite-temperature and interaction effects on spin transport constitute an interesting future direction (see also Refs.~\cite{PhysRevB.79.235321,PhysRevLett.102.076602,PhysRevLett.102.096806} for quantum point contacts). Other intriguing future directions would be to study the details of the tunnel-coupling between a TI edge and a ferromagnetic  contact~\cite{PhysRevB.89.035408,PhysRevLett.117.166806,sayed2016multi,2020PhLA..38426228D} or the effects of electric fields in relatively clean systems and investigate the potential to control spin polarization electrically~\cite{PhysRevB.104.115205}.

\section*{Acknowledgments}
We thank Yuli Lyanda-Geller, Pramey Upadhyaya, and Igor \v{Z}uti\'{c} for valuable discussions. 
J.C. would like to thankfully acknowledge the Office of Undergraduate Research at Purdue University for financial support. 
This material is based upon work supported by the U.S. Department of Energy, Office of Science, National Quantum Information Science Research Centers, Quantum Science Center. 

\bibliography{refs}

\appendix
\section{\label{Trans/Refl derivation}Derivation of transmission and reflection coefficients}

Here we derive the transmission $t$ and reflection $r$ coefficients given by Eqs.~(\ref{Transmission coefficient})--(\ref{Reflection coefficient}). We first find the eigenstates of Eq.~(\ref{Effective Hamiltonian disorder}) by rearranging the Schrödinger equation $\mathcal{H}(x)\psi = E\psi$ into a more convenient form,
\begin{equation}
    \partial_x\psi = \frac{1}{\hbar v}\left[m(x)\sigma_y + i(E+\mu)\sigma_z\right]\psi \,,
    \label{Schrodinger equation}
\end{equation}
which can then be solved through the use of a matrix exponential:
\begin{equation}
    \psi(x_0) = \exp{\eta_m\sigma_y + i\xi\sigma_z}\psi(0) \,,
    \label{Matrix exponential}
\end{equation}
where $\eta_m = \int_0^{x_0}m(x)dx/(\hbar v)$ and $\xi = (E+\mu)x_0/(\hbar v)$. Thus, by taking $0/x_0$ to be at the left/right edges of the disordered region and expanding the matrix exponential in Eq.~(\ref{Matrix exponential}), we can calculate how the disorder scatters an incoming mode. Defining $\chi = \sqrt{\xi^2 - \eta_m^2}$, the matrix exponential in Eq.~(\ref{Matrix exponential}) is equal to the scattering operator
\begin{equation}
    \hat{S} = \cos{\chi} + \frac{\eta_m\sigma_y + i\xi\sigma_z}{\chi}\sin{\chi} \,.
    \label{Scattering operator}
\end{equation}

To calculate the transmission/reflection coefficient of an incoming right-mover, we apply $\hat{S}$ to the state $\psi(0) = \ket{\uparrow} + r\ket{\downarrow}$, where $r$ is the reflection amplitude yet to be determined:
\begin{equation}
    \begin{split}
        \psi(x_0) &= \hat{S}\psi(0) \\
        &= \left[\cos{\chi} + i\frac{\xi}{\chi}\sin{\chi} - ir\frac{\eta_m}{\chi}\sin{\chi}\right]\ket{\uparrow} \\
        &+ \left[r\cos{\chi} - ir\frac{\xi}{\chi}\sin{\chi} + i\frac{\eta_m}{\chi}\sin{\chi}\right]\ket{\downarrow} \,.
    \end{split}
    \label{Right-mover scattering}
\end{equation}
Since the spin-down state on the right side of the barrier is an incoming left-mover, we know its coefficient must be zero. Hence, solving for $r$ and plugging the result into the spin-up coefficient for $t$ gives
\begin{align}
    t &= \frac{\chi^2\cos{\chi} + i\xi\chi\sin{\chi}}{\xi^2 - \eta_m^2\cos^2{\chi}} \label{Appendix trans amplitude} \,, \\
    r &= \frac{\eta_m\xi\sin^2{\chi} - i\eta_m\chi\sin{\chi}\cos{\chi}}{\xi^2 - \eta^2_m\cos^2\chi} \label{Appendix refl amplitude} \,.
\end{align}
Finally, we note that a similar analysis using an incoming left-mover gives the same coefficients, resulting in a unitary scattering matrix as given by Eq.~(\ref{Scattering matrix}) in the low-energy limit. Furthermore, the square magnitudes $\abs{t}^2$ and $\abs{r}^2$ are (restoring $\chi = \sqrt{\xi^2 - \eta_m^2}$)
\begin{align}
    \abs{t}^2 &= \frac{\xi^2 - \eta_m^2}{\xi^2 - \eta_m^2\cos^2{\sqrt{\xi^2 - \eta_m^2}}} \label{Squared trans amplitude} \,, \\
    \abs{r}^2 &= \frac{\eta_m^2\sin^2{\sqrt{\xi^2 - \eta_m^2}}}{\xi^2 - \eta_m^2\cos^2{\sqrt{\xi^2 - \eta_m^2}}} \label{Squared refl amplitude} \,.
\end{align}
Taking the scattering state near the Dirac point, ${\xi \ll \eta_m}$, these expressions are used in Eqs.~(\ref{Magnetic disorder left current})--(\ref{Magnetic disorder right current}) to calculate the spin current on the left and right of the disordered edge segment. 

\section{\label{TB model sec}Tight-binding model} 

(Note: in this Appendix we denote $\mathbf{z}$ the axis perpendicular to the monolayer, while the spin quantization axis, denoted here $\mathbf{z}'$, is tilted with respect to the normal (see the end of this section). In the main text we drop the prime from $\mathbf{z}'$ for brevity.)
Here we reproduce the tight-binding model introduced in Ref. \cite{Lau2019} and detail the disorder terms used in Sec. \ref{Results Sec}. WTe$_2$ in the $1T'$ configuration consists of a square lattice with six atoms per unit cell. The effective tight-binding model introduced by Ref. \cite{Lau2019} reduces this to a four-site square lattice, with two $d_{x^2-y^2}$ (W) orbitals and two $p_x$ (Te) orbitals per cell. 

 Adopting the notation  of Ref.~\cite{Lau2019}, we define the Pauli matrices $s_i$, $\tau_i$, and $\sigma_i$ to act on the spin, sublattice, and orbital degrees of freedom, respectively, and define the $\Gamma_i$ matrices by
\begin{align}
    \Gamma_0 &= \tau_0\sigma_0 \label{Gamma0} \,, \\
    \Gamma^\pm_1 &= \frac{1}{2}\tau_0(\sigma_0\pm\sigma_3) \label{Gamma1} \,, \\
    \Gamma^\pm_2 &= \frac{1}{4}(\tau_1+i\tau_2)(\sigma_0\pm\sigma_3) \label{Gamma2} \,, \\
    \Gamma_3 &= \frac{i}{2}(\tau_1+i\tau_2)\sigma_2 \label{Gamma3} \,, \\
    \Gamma^\pm_4 &= \frac{1}{4}(\tau_0\pm\tau_3)(\sigma_1+i\sigma_2) \label{Gamma4} \,, \\
    \Gamma^\pm_5 &= \frac{1}{2}\tau_3(\sigma_0\pm\sigma_3) \label{Gamma5} \,, \\
    \Gamma_6 &= \frac{1}{2}(\tau_1+i\tau_2)\sigma_1 \label{Gamma6} \,.
\end{align}
To be explicit, a term of the form $s_i\tau_j\sigma_k$ acts on the operator
\begin{equation}
    \begin{split}
        c_{\Vec{r}} = [ &c_{\Vec{r}\,\uparrow Ad},\, c_{\Vec{r}\,\uparrow Ap},\, c_{\Vec{r}\,\uparrow Bd},\, c_{\Vec{r}\,\uparrow Bp}, \\
        &c_{\Vec{r}\,\downarrow Ad},\, c_{\Vec{r}\,\downarrow Ap},\, c_{\Vec{r}\,\downarrow Bd},\, c_{\Vec{r}\,\downarrow Bp} ] \,,
    \end{split}
    \label{c operator}
\end{equation}
where $c_{\Vec{r}slo}$ annihilates a spin $s\in\{\uparrow,\downarrow\}$ electron on sublattice $l\in\{A,B\}$ and orbital $o\in\{d,p\}$.

With these definitions, the tight-binding Hamiltonian can be written as $H=\sum_{\Vec{r}}\left[H_0(\Vec{r}) + \lambda(\Vec{r})H_{\mathrm{SOC}}(\Vec{r}) + \delta H(\Vec{r})\right]$, where \cite{Lau2019}
\begin{equation}
    \begin{split}
        H_0(\Vec{r}) &=
        \frac{\mu_d}{2}c^\dagger_{\Vec{r}}\,\Gamma^+_1c_{\Vec{r}} + \frac{\mu_p}{2}c^\dagger_{\Vec{r}}\,\Gamma^-_1c_{\Vec{r}} \\
        &+ \frac{t_{dx}}{2}c^\dagger_{\Vec{r}}\,\Gamma^+_1(c_{\Vec{r}+\Vec{a}}+c_{\Vec{r}-\Vec{a}}) \\
        &+ \frac{t_{px}}{2}c^\dagger_{\Vec{r}}\,\Gamma^-_1(c_{\Vec{r}+\Vec{a}}+c_{\Vec{r}-\Vec{a}}) \\
        &+ \frac{t_{py}}{2}c^\dagger_{\Vec{r}}\,\Gamma^-_1(c_{\Vec{r}+\Vec{b}}+c_{\Vec{r}-\Vec{b}}) \\
        &+ t_{dAB}\,c^\dagger_{\Vec{r}}\,\Gamma^+_2(c_{\Vec{r}-\Vec{b}+\Vec{\Delta}_1}+c_{\Vec{r}+\Vec{a}-\Vec{b}+\Vec{\Delta}_1}) \\
        &+ t_{pAB}\,c^\dagger_{\Vec{r}}\,\Gamma^-_2(c_{\Vec{r}+\Vec{\Delta}_2}+c_{\Vec{r}+\Vec{a}+\Vec{\Delta}_2}) \\
        &+ t_{0AB}\,c^\dagger_{\Vec{r}}\,\Gamma_3(c_{\Vec{r}+\Vec{\Delta}_3}-c_{\Vec{r}+\Vec{a}+\Vec{\Delta}_3}) \\
        &- t_{0x}\,c^\dagger_{\Vec{r}}\,\Gamma^+_4(c_{\Vec{r}+\Vec{a}+\Vec{\Delta}_4}-c_{\Vec{r}-\Vec{a}+\Vec{\Delta}_4}) \\
        &- t_{0x}\,c^\dagger_{\Vec{r}}\,\Gamma^-_4(c_{\Vec{r}+\Vec{a}-\Vec{\Delta}_4}-c_{\Vec{r}-\Vec{a}-\Vec{\Delta}_4}) \\
        &+ t_{0ABx}\,c^\dagger_{\Vec{r}}\,\Gamma_3(c_{\Vec{r}-\Vec{a}+\Vec{\Delta}_3}-c_{\Vec{r}+2\Vec{a}+\Vec{\Delta}_3}) \\
        &+ \mathrm{H.c.} \,,
    \end{split}
    \label{TB Hamiltonian H0}
\end{equation}
and
\begin{equation}
    \begin{split}
        H_{\mathrm{SOC}}(\Vec{r}) &=
        - \frac{i}{2}c^\dagger_{\Vec{r}}\,(\lambda^z_{dx}s_z + \lambda^y_{dx}s_y)\Gamma^+_5(c_{\Vec{r}+\Vec{a}}-c_{\Vec{r}-\Vec{a}})\\
        &- \frac{i}{2}c^\dagger_{\Vec{r}}\,(\lambda^z_{px}s_z + \lambda^y_{px}s_y)\Gamma^-_5(c_{\Vec{r}+\Vec{a}}-c_{\Vec{r}-\Vec{a}}) \\
        &- i\lambda^y_{0AB}c^\dagger_{\Vec{r}}\,s_y\Gamma_6(c_{\Vec{r}+\Vec{\Delta}_3} + c_{\Vec{r}+\Vec{a}+\Vec{\Delta}_3}) \\
        &- ic^\dagger_{\Vec{r}}\,(\lambda^z_0s_z + \lambda^y_0s_y)\Gamma^+_4c_{\Vec{r}+\Vec{\Delta}_4} \\
        &+ ic^\dagger_{\Vec{r}}\,(\lambda^z_0s_z + \lambda^y_0s_y)\Gamma^-_4c_{\Vec{r}-\Vec{\Delta}_4} \\
        &- ic^\dagger_{\Vec{r}}(\lambda'^z_0s_z + \lambda'^y_0s_y)\Gamma^+_4c_{\Vec{r}-\Vec{b}+\Vec{\Delta}_4} \\
        &+ ic^\dagger_{\Vec{r}}(\lambda'^z_0s_z + \lambda'^y_0s_y)\Gamma^-_4c_{\Vec{r}+\Vec{b}-\Vec{\Delta}_4} \\
        &+ \mathrm{H.c.} \,,
    \end{split}
    \label{TB Hamiltonain SOC}
\end{equation}
are the spin-rotation symmetric and spin-orbit coupling terms, respectively, $\lambda(\Vec{r})$ is a potentially site-dependent scale factor modifying the SOC strength ($\lambda(\Vec{r})=1$ for physical WTe$_2$ and $=0$ for no SOC), and $\delta H(\Vec{r})$ describes any additional disorder terms. The parameter values used in Eqs.~(\ref{TB Hamiltonian H0})--(\ref{TB Hamiltonian H0}) can be found in Table~\ref{TB parameters}.

\begingroup
\setlength{\tabcolsep}{6pt} 
\renewcommand{\arraystretch}{1} 
\begin{table}
    \centering
    \begin{tabular}{|cc|cc|}
        \hline
        \multicolumn{2}{|c|}{$H_0$ (eV)} & \multicolumn{2}{|c|}{$H_{\mathrm{SOC}}$ (meV)} \\
        \hline
        $\mu_d$ & 0.24 & $\lambda^z_{dx}$ & -8 \\
        $\mu_p$ & -2.25 & $\lambda^y_{dx}$ & -31 \\
        $t_{dx}$ & -0.41 & $\lambda^z_{px}$ & -10 \\
        $t_{px}$ & 1.13 & $\lambda^y_{px}$ & -40 \\
        $t_{py}$ & 0.13 & $\lambda^y_{0AB}$ & 11 \\
        $t_{dAB}$ & 0.51 & $\lambda^z_0$ & 12 \\
        $t_{pAB}$ & 0.4 & $\lambda^y_0$ & 51 \\
        $t_{0AB}$ & 0.39 & $\lambda'^z_0$ & 12 \\
        $t_{0x}$ & 0.14 & $\lambda'^y_0$ & 50 \\
        $t_{0ABx}$ & 0.29 & & \\
        \hline
    \end{tabular}
    \caption{Parameter values used in the tight-binding model of 1T'-WTe$_2$ \cite{Lau2019} (see Eqs.~(\ref{TB Hamiltonian H0})--(\ref{TB Hamiltonain SOC})).}
    \label{TB parameters}
\end{table}
\endgroup

\begin{figure}
    \centering
    \includegraphics[width=.48\textwidth]{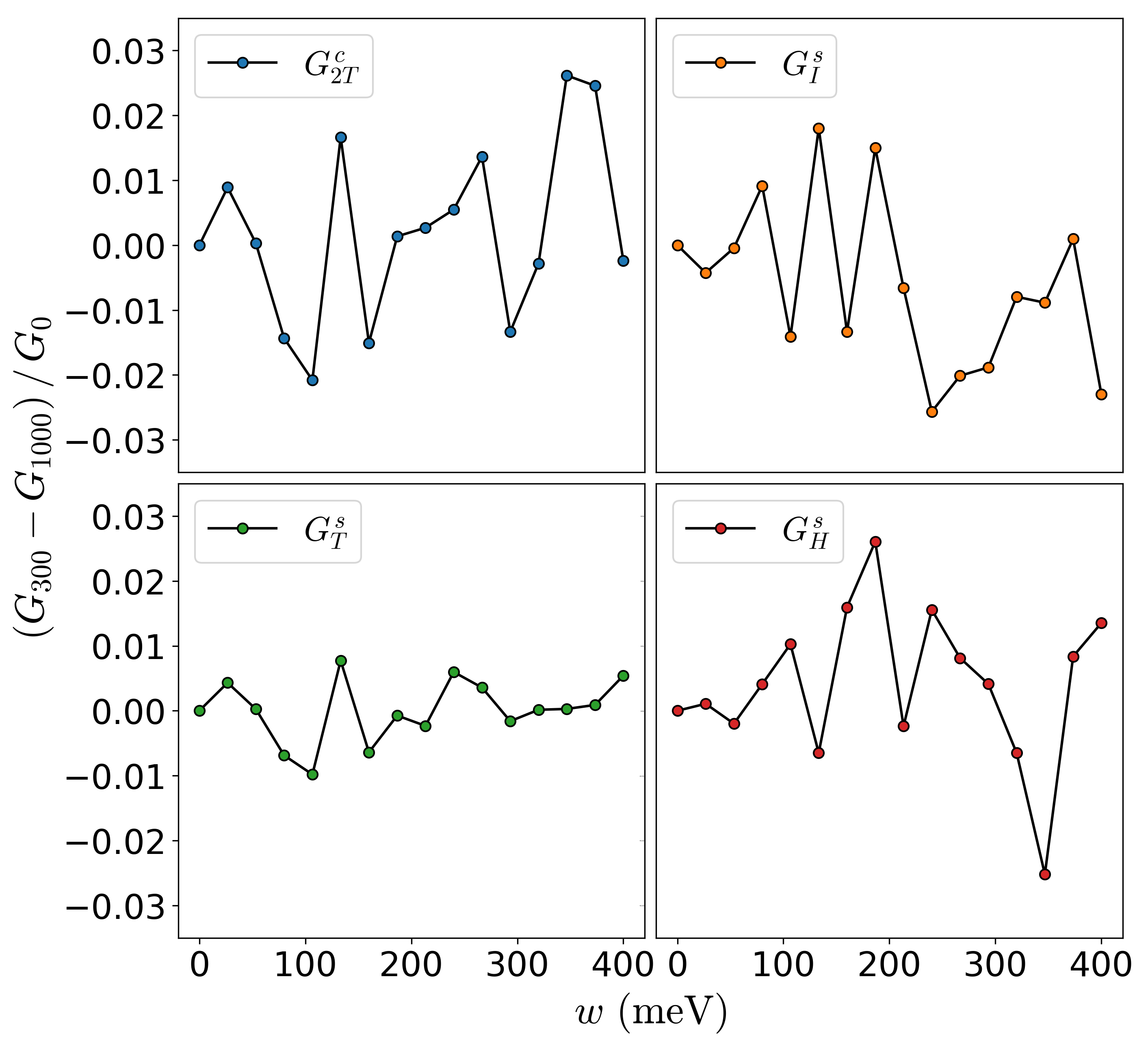}
    \caption{Normalized differences between 300 sample and 1000 sample averages. The disorder term is a localized magnetic perturbation as discussed in Sec.~\ref{Magnetic disorder sec} and shown in Fig.~\ref{Magnetic disorder plots}b.}
    \label{Diff_between_avgs}
\end{figure}

\begin{figure}
    \centering
    \includegraphics[width=.48\textwidth]{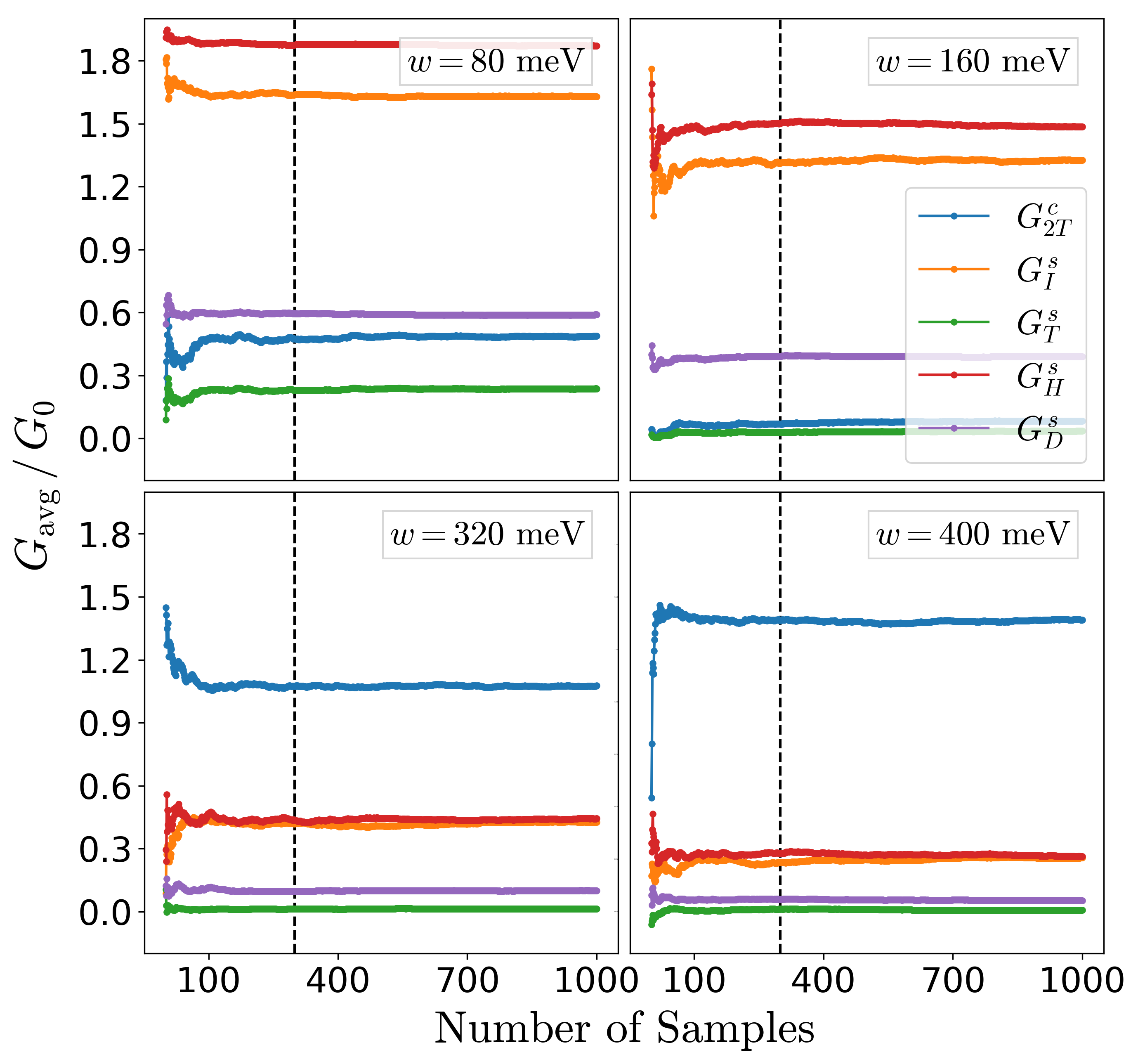}
    \caption{Disorder-averaged conductance components versus the number of samples used at fixed $w$. The disorder term is a localized magnetic perturbation as discussed in Sec.~\ref{Magnetic disorder sec} and shown in Fig.~\ref{Magnetic disorder plots}b.}
    \label{convergence_plot}
\end{figure}

In Sec. \ref{Results Sec} we study the effects of several on-site and hopping disorder terms. In Sec. \ref{Scalar disorder sec} we study on-site scalar disorder terms of the form $\delta H(\Vec{r}) = u(\Vec{r})c^\dagger_{\Vec{r}}\,s_0\Gamma_0c_{\Vec{r}}$, where $u(\Vec{r})$ is drawn from a Gaussian of mean 0 and standard deviation $w$. We also study spin-conserving disorder in the SOC strength by having $\lambda(\Vec{r})$ be drawn from a Gaussian of mean 1 and standard deviation $\delta\lambda$. In Sec.~\ref{Time-reversal symmetric  disorder sec} we 
break spin-rotational symmetry (while 
preserving TR symmetry)  by adding a disorder term $\delta H(\Vec{r}) = i\lambda'^x_0(\Vec{r})c^\dagger_{\Vec{r}}s_x(\Gamma^+_4c_{\Vec{r}-\Vec{b}+\Vec{\Delta}_4}-\Gamma^-_4c_{\Vec{r}+\Vec{b}-\Vec{\Delta}_4}) + \mathrm{H.c.}$, with $\lambda'^x_0(\Vec{r})$ drawn from a Gaussian of mean 0 and standard deviation $w$. Finally, in Sec. \ref{Magnetic disorder sec} we break both TR and spin-rotational symmetry by including an on-site perturbation $\delta H(\Vec{r})=m(\Vec{r})c^\dagger_{\Vec{r}}\,s_x\Gamma_0c_{\Vec{r}}$, with $m(\Vec{r})$ once again drawn from a Gaussian of mean 0 and standard deviation $w$.

Finally, we comment on the edge state spin quantization axis of a pristine WTe$_2$ obtained from Eqs.~(\ref{TB Hamiltonian H0})--(\ref{TB Hamiltonain SOC}); let us denote the axis $\mathbf{z}'$ in this section. Noting the lack of an $s_x$ term in Eq.~(\ref{TB Hamiltonain SOC}), it is clear that the spin quantization axis  $\mathbf{z}'$ lies in the $yz$-plane. Numerically, we find $\mathbf{z}' \approx \mathbf{z} \cos\theta + \mathbf{y} \sin\theta$ with $\theta \approx 76.7^\circ$ and measure the spin current using Eq.~(\ref{Landauer-Buttiker formula}) along this axis. Furthermore, as detailed in the preceding paragraph, the  spin-symmetry breaking disorder terms we consider are $\mathbf{x}$-polarized and thus perpendicular to both $\mathbf{z}$ and $\mathbf{z}'$, ensuring that these perturbations fully break the spin-rotational symmetry. In the main text, including Eq.~(\ref{Effective Hamiltonian}), we drop the prime from $\mathbf{z}'$ and simply denote the spin quantization axis $\mathbf{z}$.

\section{\label{Convergence sec}Convergence of disorder-averaged conductance}

Here we confirm the convergence of the disorder-averaged conductance components in the presence of magnetic disorder. To do this, we have extended our calculations for Fig.~\ref{Magnetic disorder plots}b to include 1000 samples (in comparison to the 300 samples used in the plot). We display the results of these calculations in Figs.~\ref{Diff_between_avgs}--\ref{convergence_plot}. In Fig.~\ref{Diff_between_avgs} we plot difference in the conductance values averaged over 300 and 1000 samples, normalized by their corresponding conductance quanta $G_0$ ($e^2/h$ for charge conductance and $e/(4\pi)$ for spin conductance). The difference between these averages is less than $\pm0.03\,G_0$ for each component, which is small enough for our purposes. Meanwhile, in Fig.~\ref{convergence_plot}, we plot the average conductance values versus the number of samples for fixed values of the disorder strength $w$. We note the averages appear to converge to their long-run values after a few hundred samples, with most of the fluctuations occurring well before 300 samples (marked by a dashed line).

\end{document}